\documentclass{article}

\usepackage[letterpaper,top=2cm,bottom=2cm,left=3cm,right=3cm,marginparwidth=1.75cm]{geometry}
\usepackage{authblk}
\usepackage{url}

\usepackage{amsmath}
\usepackage{mathtools}

\usepackage{graphicx}
\usepackage{subcaption}

\usepackage{multirow}
\usepackage{tabularx}
\newcolumntype{C}{>{\centering\arraybackslash}X}

\usepackage{tikz}
\let\push\undefined
\usetikzlibrary{quantikz2}

\title{
Quantum-classical hybrid models based on error correction for time series forecasting
}

\author[1,*]{Jonathan H. A. de Carvalho}
\author[2]{Filipe C. de L. Duarte}
\author[1]{Fernando M. de Paula Neto}
\author[1]{Paulo S. G. de Mattos Neto}

\affil[1]{Centro de Informática, Universidade Federal de Pernambuco, Recife, Brazil}
\affil[2]{Departamento de Finanças e Contabilidade, Universidade Federal da Paraíba, João Pessoa, Brazil}
\affil[*]{Corresponding author. E-mail: \textit{jhac@cin.ufpe.br}}

\date{}

\begin{document}

\maketitle

\begin{abstract}
Time series forecasting largely benefits from combining the strengths of different models, especially using a scheme where a model corrects another model by capturing supplementary patterns from forecasting errors. Concurrently, quantum models are providing a means to augment the classical capacity, including in time series forecasting, by acting alongside classical models in hybrid architectures. In this work, we propose the first forecasting system based on error correction that jointly uses quantum and classical models. Here, quantum models first extract patterns by exploring quantum phenomena, and classical models capture the remaining patterns from the quantum errors. Compared to classical single models and classical-classical hybrid models based on error correction, the complementary capacity that emerges from this quantum-classical system provided the best results in most of the addressed problems. Therefore, this work paves the way to introduce quantum models in established hybridization schemes for time series forecasting.
\end{abstract}
\section{Introduction}

Machine Learning is characterized by the exploration of algorithms to automatically discover patterns from the data of a given task to support subsequent actions~\cite{bishop_ml-book}. In time series forecasting, the idea is to model patterns from the past observations of a phenomenon in order to predict future values~\cite{hyndman_tsf-book}. The stock market~\cite{bao_stock-tsf-review} and the sectors that supply electricity~\cite{ugbehe_elec-tsf-review} and water~\cite{boo_water-tsf-review} are among the practical applications in which forecasting is widely required.

Time series forecasting was initially addressed with statistical models, although these models focus on linear patterns in data with stable mean and variance~\cite{box&jenkins_ts-book}. Subsequently, machine learning models were introduced into the field to uncover relationships directly from the input data, including non-linear time patterns~\cite{masini_ml-tsf-review}. Deep learning models, in turn, offer a better capacity to capture complex non-linear patterns, especially local and long-term temporal dependencies~\cite{kong_dl-tsf-review}. Nevertheless, challenging forecasting problems can be significantly better predicted with simple linear models than with sophisticated deep models, as the second is prone to overfitting~\cite{zeng_formers-vs-linear-tsf}.

To guarantee accurate performance in forecasting practice, the complementary strengths of distinct models are frequently integrated through several hybrid structures~\cite{hajirahimi_hybrid-tsf-review}, including advanced schemes that further combine existing hybrid structures~\cite{hajirahimi-hybrid-hybrid-tsf-review}. Hybridization is a long-standing theory that allows different individuals to capture different information in forecasting tasks~\cite{clemen_tsf-hybrid-theory-review}. In particular, the error correction method provided a simple yet effective hybridization scheme in which different time patterns can be modeled sequentially.

Hybridization through error correction was first illustrated by using a Multilayer Perceptron~(MLP) to capture patterns from the forecasting errors of a statistical model~\cite{zhang_err-corr-hybrid-tsf}. In this way, the MLP model was able to predict errors that were used to correct the predictions of the statistical model. Statistical models can also be corrected by deep learning models such as Neural Basis Expansion Analysis for Time Series Forecasting~(NBEATS)~\cite{duarte_arima-nbeats}. Further combinations include correcting a Support Vector Regression using a Convolutional Neural Network~(CNN)~\cite{zhu_svr-tcn}, as well as correcting a Long Short-Term Memory~(LSTM) using another LSTM~\cite{ma_lstm-lstm} or a convolutional LSTM~\cite{melalkia_lstm-convlstm}. Additionally, a CNN with an attention mechanism and a final Gated Recurrent Unit can correct an Autoformer-based model~\cite{cheng_autoformer-cnnattgru}.

Quantum Machine Learning, which can explore quantum phenomena such as superposition and entanglement~\cite{nielsen_qc-book}, is expected to provide models that offer practical advantages over classical models in multiple applications~\cite{cerezo_qml-vqc-review, zhao_large-data-q-efficiency}. The quantum advantage relies primarily on the ability to perform nonlinear mappings of classical data to exponentially large quantum spaces. Data analysis can then be facilitated in quantum spaces that are hard to simulate classically~\cite{havlicek_q-feat-spaces, schuld_q-feat-spaces}. The quantum encoding of classical data is a remarkable operation that not only captures data nonlinearities, but also defines the functions that can be expressed. Such an encoding ultimately enables quantum models to learn any continuous function~\cite{salinas_universal-vqc, schuld_encoding-expressive-power}.

Compared to classical models, quantum models can produce competitive results with fewer trainable parameters by leveraging the entanglement phenomenon to capture data correlations~\cite{schuld_entangling-vqc}. For an equal number of parameters instead, quantum models can express more functions than classical models and converge faster to lower errors~\cite{abbas_vqc-power}. Furthermore, the number of training data points required for quantum models to accurately predict unseen data scales efficiently with the number of parameters~\cite{caro_vqc-generalization}.

Despite these favorable properties, quantum models struggle to achieve an advantage in real-world applications. This difficulty originates from a flattening phenomenon that occurs in the optimization landscapes of such models. Gradients disappear in flat landscapes, and thus quantum models cannot be properly adjusted for the learning problem~\cite{larocca_bp-review}. However, rather than applying quantum models to seek a quantum advantage, quantum components can be connected to classical components, forming hybrid architectures that augment the classical capacities~\cite{schuld_q-adv-alternatives, callison_hybrid-qc-algos}. As an example, the internal operations of an LSTM can be coupled with quantum models to extract features and compress data while exploring the entanglement property~\cite{chen_vqc-lstm-ops}, which has proven effective in time series forecasting.

Quantum-enhanced LSTM-based models for time series forecasting still integrate quantum models with classical Fully Connected~(FC) layers in internal operations~\cite{yu_fc-vqc-lstm, cao_linear-vqc-lstm}. The quantum-enhanced LSTM can even be preceded by a CNN~\cite{naz_cnn-vqc-lstm} and succeeded by a Random Forest regressor~\cite{dong_cnn-vqc-lstm-rfr}. Other quantum-classical architectures for forecasting tasks include the application of conventional LSTM models followed by an FC layer and a quantum model at the end of the pipeline~\cite{ceschini_lstm-fc-vqc}, a quantum model between two FC layers~\cite{ruiz_fc-vqc-fc}, and arrangements such as LSTM-FC-Q-FC~\cite{hong_lstm-fc-vqc-fc} and CNN-LSTM-Q-FC~\cite{sun_cnn-lstm-vqc-fc}, where ``Q'' refers to a quantum model. Ultimately, quantum-classical architectures are designed to take advantage of quantum properties in order to enhance the classical capacity to capture time patterns in data.

In this work, we proceed on the premise that time series problems contain patterns that can only be modeled by exploring quantum phenomena, whereas other patterns can be regularly modeled by exploring the classical strengths. Thus, a better forecasting system could be developed if a quantum model is used first to capture time patterns in the quantum space, and then a classical model is used to capture the remaining time patterns. To investigate this assumption, we propose the first exploration of quantum models, along with classical models, in the established error correction scheme for time series forecasting.

Here, quantum models learn patterns in the given time series, whereas classical models are responsible for capturing patterns in the quantum forecasting errors. The experiments conducted in this study show that quantum-classical models based on error correction can produce better results than both classical single models and classical-classical models based on the same hybridization scheme. Therefore, a sequential scheme composed of quantum and classical models is able to handle data relationships in a quantum-classical complementary manner that uncovers a novel capacity for time series forecasting tasks.
\section{Results}

\subsection{Error-correction-based quantum-classical models}

Here, each quantum-classical integration comprises: (1) a quantum model that captures patterns from the past values of a given time series and (2) a classical model that captures patterns from past quantum errors. Quantum forecasting errors are defined as the difference between the quantum predictions and the original observations. Then, the next value in the time series can be first approximated by the quantum model, while the classical model predicts the quantum error. The outputs of these two models are combined via an addition operation, resulting in a hybrid and better prediction that is a classical refinement of the first quantum approximation. The supplementary material covers a mathematical definition and also illustrates this quantum-classical hybridization based on error correction.

However, this hybrid formulation does not always improve the classical performance in forecasting tasks for any quantum-classical combination. In other words, complementary pattern learning that goes beyond the classical capacity does not arise consistently in practice simply because quantum and classical models are jointly used. Instead, the classical performance is boosted with only particular quantum-classical pairings. Later in this work, a systematic strategy for selecting quantum and classical components that effectively cooperate with each other is delineated.

\begin{figure}[ht]
    \centering
    \includegraphics[width=0.7\textwidth]{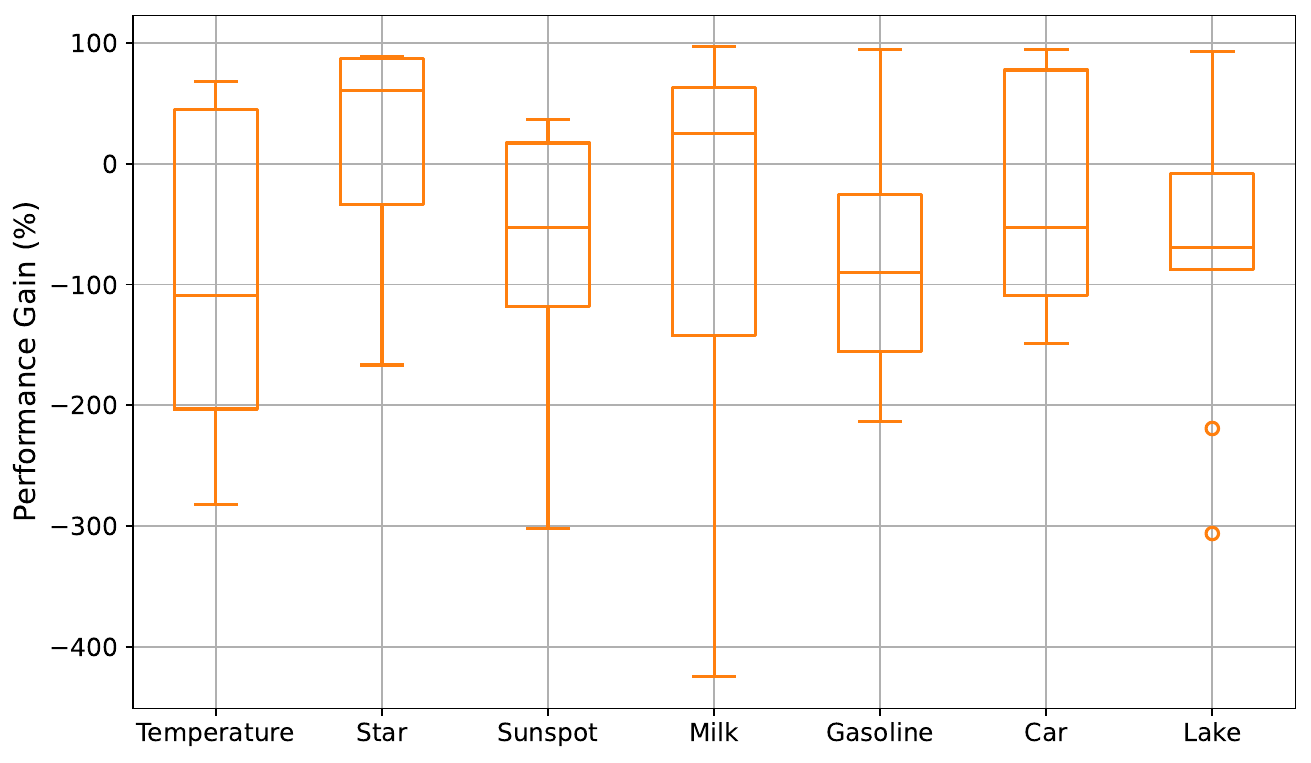}
    \caption{Percentage performance gains that quantum-classical combinations based on error correction provided to the base classical models in different forecasting problems. In each of the seven problems, a box gathers nine samples of performance gains, which are obtained combining three quantum models with three classical models and then comparing with the respective classical models in the given problem.}
    \label{fig:hybrid-gains}
\end{figure}

For now, we scrutinize the effectiveness of error-correction-based quantum-classical models in leading to better performances from the perspective of the classical models involved in the combinations. To be effective, a quantum-classical model needs to properly capture complementary time patterns that the classical model cannot capture alone. An improvement can be measured by the performance gain that the combination $\text{Q}_i+\text{C}_j$ leads to the corresponding model $\text{C}_j$ in unseen data, as follows:

\begin{equation*}
    \frac{P_{\text{C}_j} - P_{\text{Q}_i+\text{C}_j}}{P_{\text{C}_j}} \times 100,
\end{equation*}

\noindent
where each $P_M$ refers to the performance of the model $M$. The metric used to calculate these performances is the Mean Squared Error (MSE), which is discussed in the supplementary material. Three quantum models and three classical models were applied in seven forecasting problems, providing nine quantum-classical combinations to compare with the corresponding classical model in each problem. These quantum models, classical models, and forecasting problems are described in the Methods section, while the supplementary material details the results of each model in each problem. Here, Fig.~\ref{fig:hybrid-gains} gathers the nine performance gains obtained via the quantum-classical combinations in each problem. In fact, gains below zero actually represent losses.

As the median performance gain is really above zero in Star and Milk only, at least half of the quantum-classical combinations led to performance degradation in the other five problems. Especially in the Gasoline and Lake problems, the entire box is below zero, which means that 75\% of the combinations generated losses in terms of the classical performances. Some combinations still resulted in performance losses of approximately 300\% in Temperature, Sunspot, and Lake. In contrast to solving the problem as a single model, a classical model even produced an MSE value more than 400\% worse when combined with a quantum model in Milk.

On the other hand, in all problems, there exist combinations in which the classical models take advantage of the quantum models applied in advance, with cases of performance gains of almost 100\%. When calculating performance gains, a value close to 100\% is measured if $P_{\text{Q}_i+\text{C}_j}$ approximates zero. Therefore, the quantum and classical models in such combinations cooperated with each other to the fullest. In these cases, the data relationships handled by the quantum models complement the classical models nearly to the limit. Ultimately, this potential motivates the need for a principled mechanism to solve forecasting problems maximizing the quantum-classical cooperation. The proposed procedure to choose the appropriate quantum and classical models to combine in a problem is defined in the next section.
\subsection{An error-correction-based quantum-classical system}

\begin{figure}[ht!]
    \centering
    \includegraphics[width=0.75\textwidth]{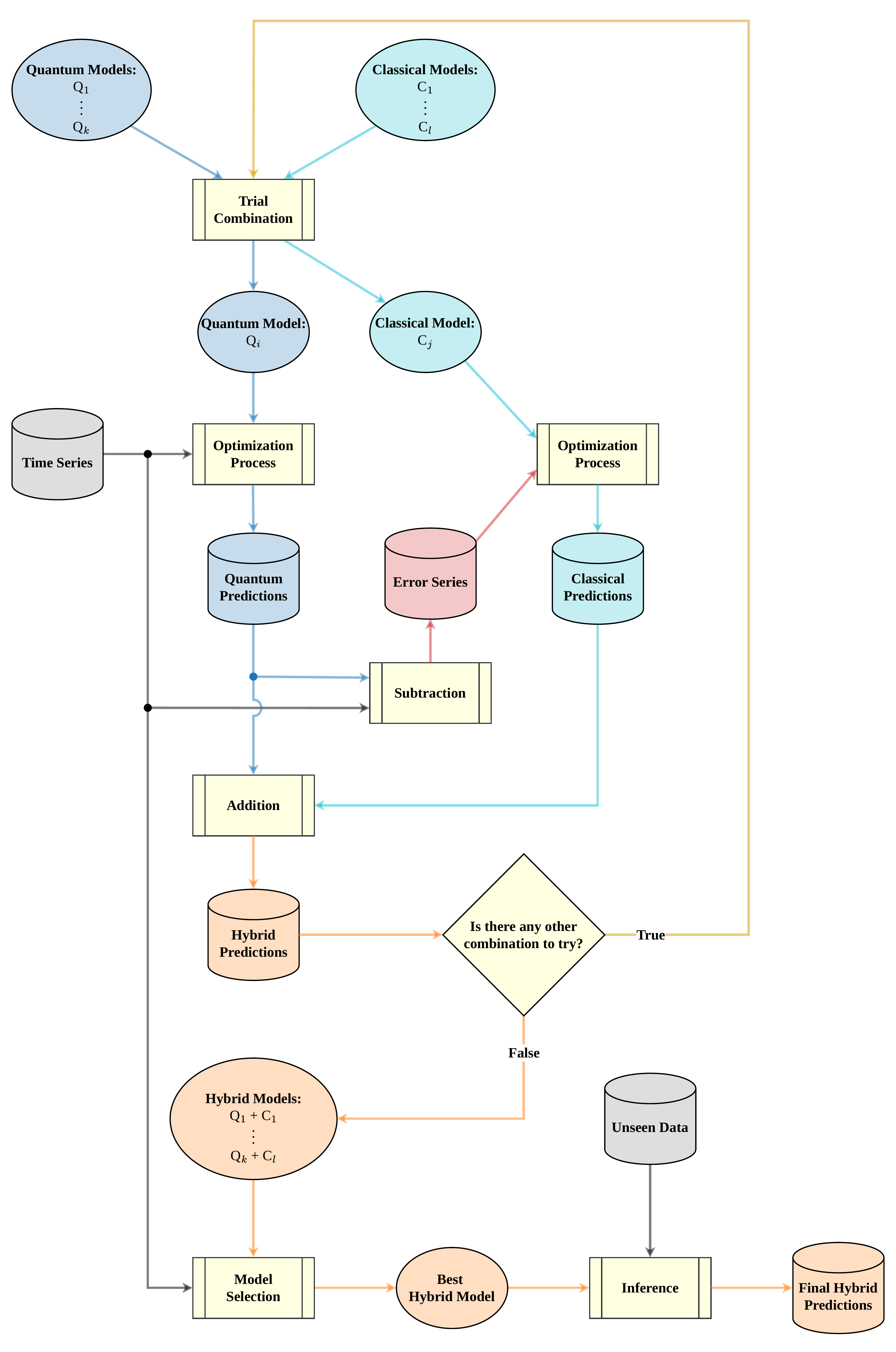}
    \caption{Proposed system to optimize the error-correction-based quantum-classical hybridization for a given forecasting task. From pools that provide different model components, the procedure tries each combination of a quantum model and a classical model. The quantum model learns and predicts the original time series, which gives errors as the subtraction of the quantum predictions from the original observations. The classical model then learns and predicts the quantum errors. Adding the classical predictions to the quantum predictions generates the hybrid predictions of the current combination of models. Once there is no other combination to try, the procedure can select the best hybrid model to be finally used when predicting unseen data.}
    \label{fig:prop-sys-diag}
\end{figure}

Depending on the participating models, an ideal complementary learning can be obtained by combining quantum and classical models via the error correction method. We therefore propose a system that explicitly tries candidate models from both paradigms and then identifies the hybrid configuration that likely yields the best performance in unseen data. Fig.~\ref{fig:prop-sys-diag} presents the proposed procedure to empirically optimize the error-correction-based quantum-classical integration of models in a problem-dependent manner.

The procedure starts with a pool of quantum models $\text{Q}_1, \cdots, \text{Q}_k$ and a pool of classical models $\text{C}_1, \cdots, \text{C}_l$. The first step is to sample a combination, which is formed by a quantum model $\text{Q}_i$ and a classical model $\text{C}_j$. As already stated, $\text{Q}_i$ learns patterns directly from the original time series. The quantum predictions can then be subtracted from the original time series, generating an error series from which $\text{C}_j$ learns patterns to produce predictions. Hybrid predictions are finally obtained by adding the classical predictions to the quantum predictions. The procedure restarts if there is any other combination of quantum and classical components to try.

Once there is no other combination to try, the system proceeds with a pool of quantum-classical models $\text{Q}_1 + \text{C}_1, \cdots, \text{Q}_k + \text{C}_l$. The next step is to select the hybrid candidate that better anticipates the original observations. This hybrid candidate is adopted as the configuration that better captured complementary time patterns and thus likely solves the problem optimally. Thus, when predicting unseen data, the selected configuration produces the predictions that represent the quantum-classical integration based on error correction in further evaluations against classical baselines.
\subsection{Augmenting the classical capacity via Q+C models}

In the same way that the quantum-classical integration can be optimized in a problem-dependent manner, classical single models and error-correction-based classical-classical models can also be empirically selected before an application in unseen data. Thus, the best quantum-classical performance in a forecasting task can be contrasted with the best performance provided by both a classical single model and two classical models applied sequentially through the error correction method. Such a comparison enables the evaluation of whether a two-step scheme composed of quantum and classical components is actually required to provide an improved performance in practice.

From three quantum models and three classical models, the system already presented in Fig.~\ref{fig:prop-sys-diag} provides the best integration of models among nine quantum-classical combinations, which will be simply named ``Q+C'' in each of the seven addressed problems. Similarly, ``C'' refers to the best classical single model chosen from the same three classical models in each problem. Finally, when applied in both steps of the error correction scheme, three classical models generated nine classical-classical combinations, from which the best hybrid baseline ``C+C'' is selected. See the supplementary material for detailed results. Fig.~\ref{fig:preds} shows the predictions of Q+C, C, and C+C for the seven problems addressed here, where the target series and the three prediction series are represented by the black, orange, cyan, and green lines, respectively, in each case.

\begin{figure}[htp!]
    \centering
    
    \begin{subfigure}{0.39\textwidth}
        \includegraphics[width=\textwidth]{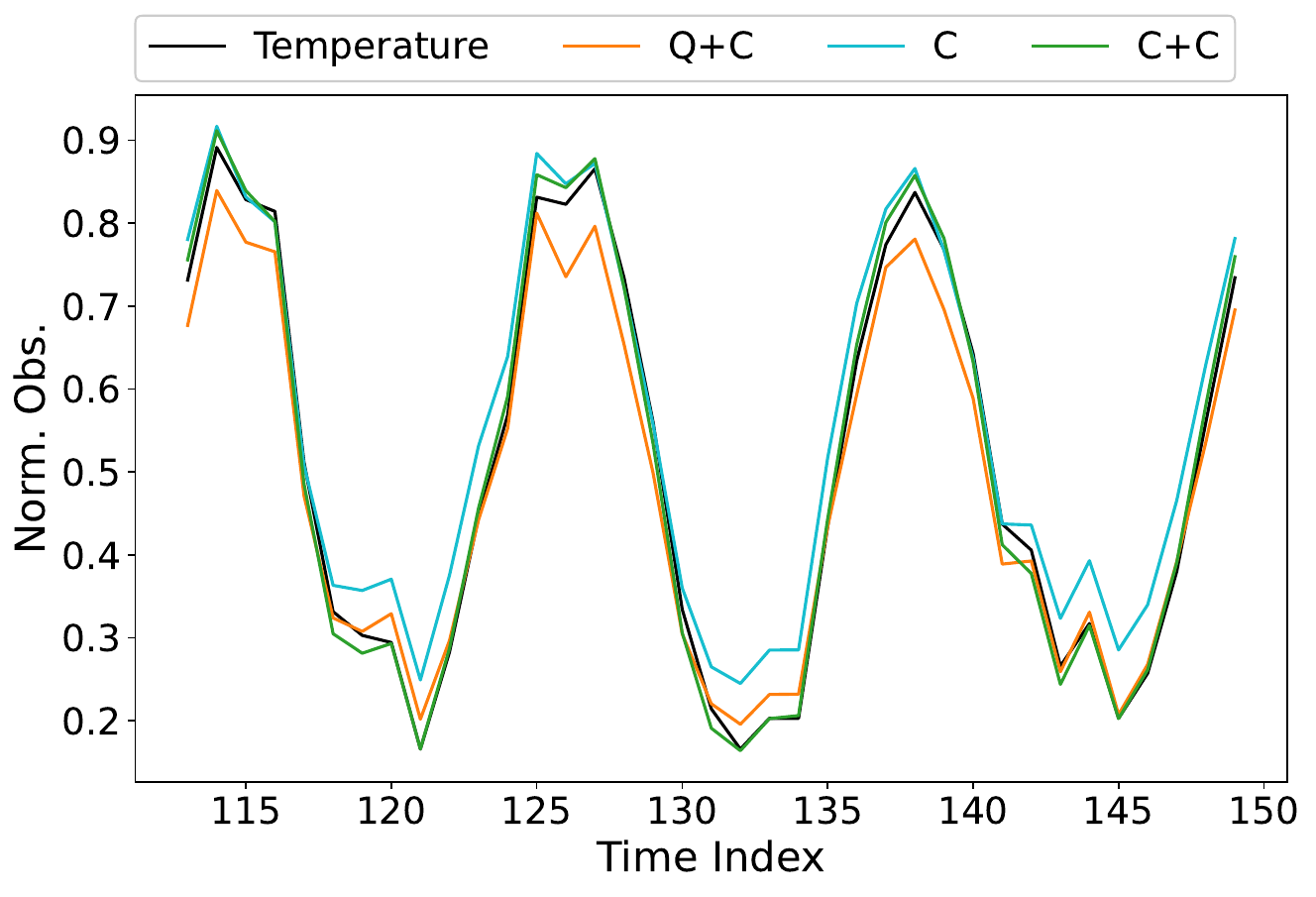}
        \caption{}
        \label{fig:temperature-preds}
    \end{subfigure}
    \hspace{1cm}
    \begin{subfigure}{0.39\textwidth}
        \includegraphics[width=\textwidth]{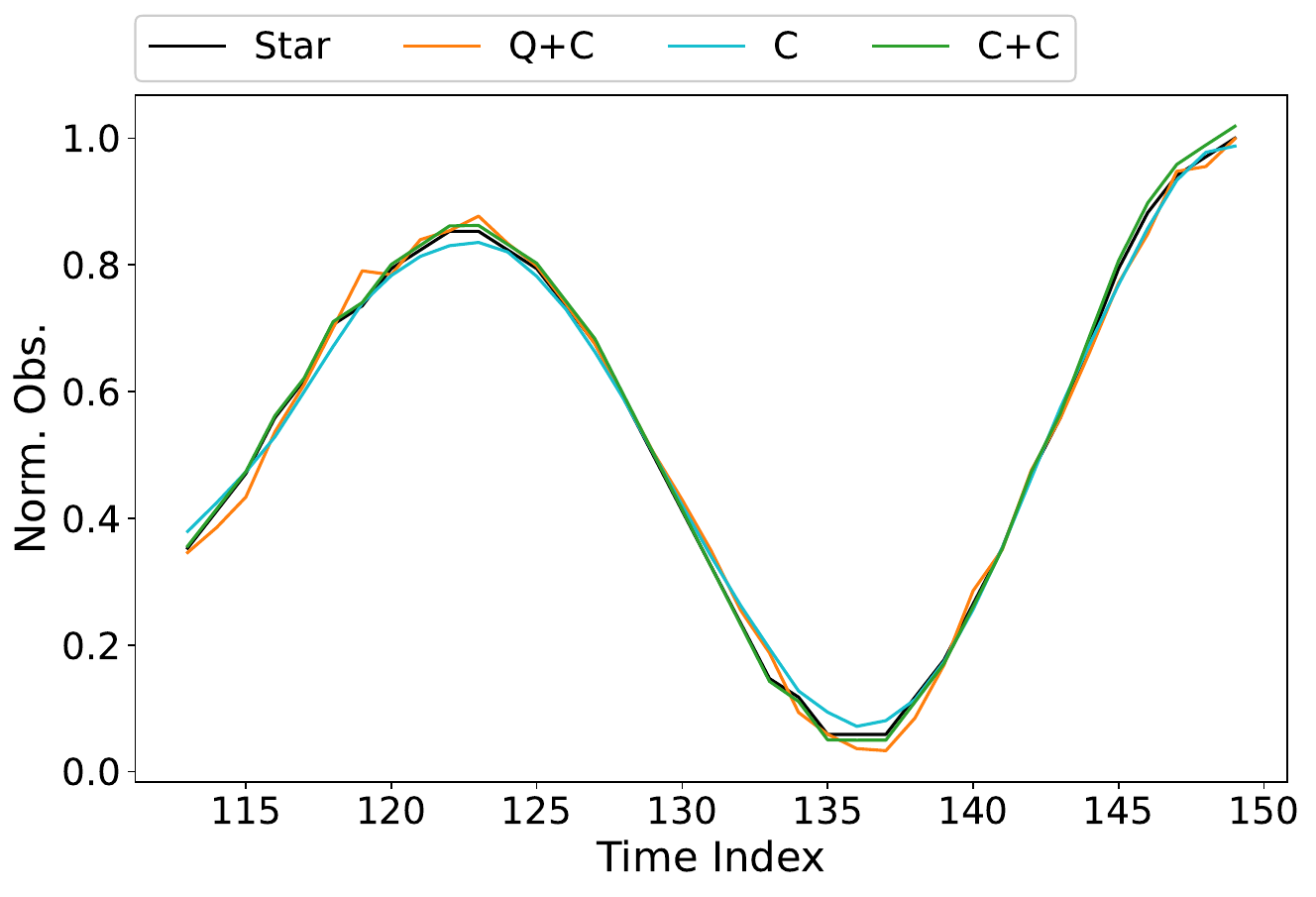}
        \caption{}
        \label{fig:star-preds}
    \end{subfigure}
    
    \begin{subfigure}{0.39\textwidth}
        \includegraphics[width=\textwidth]{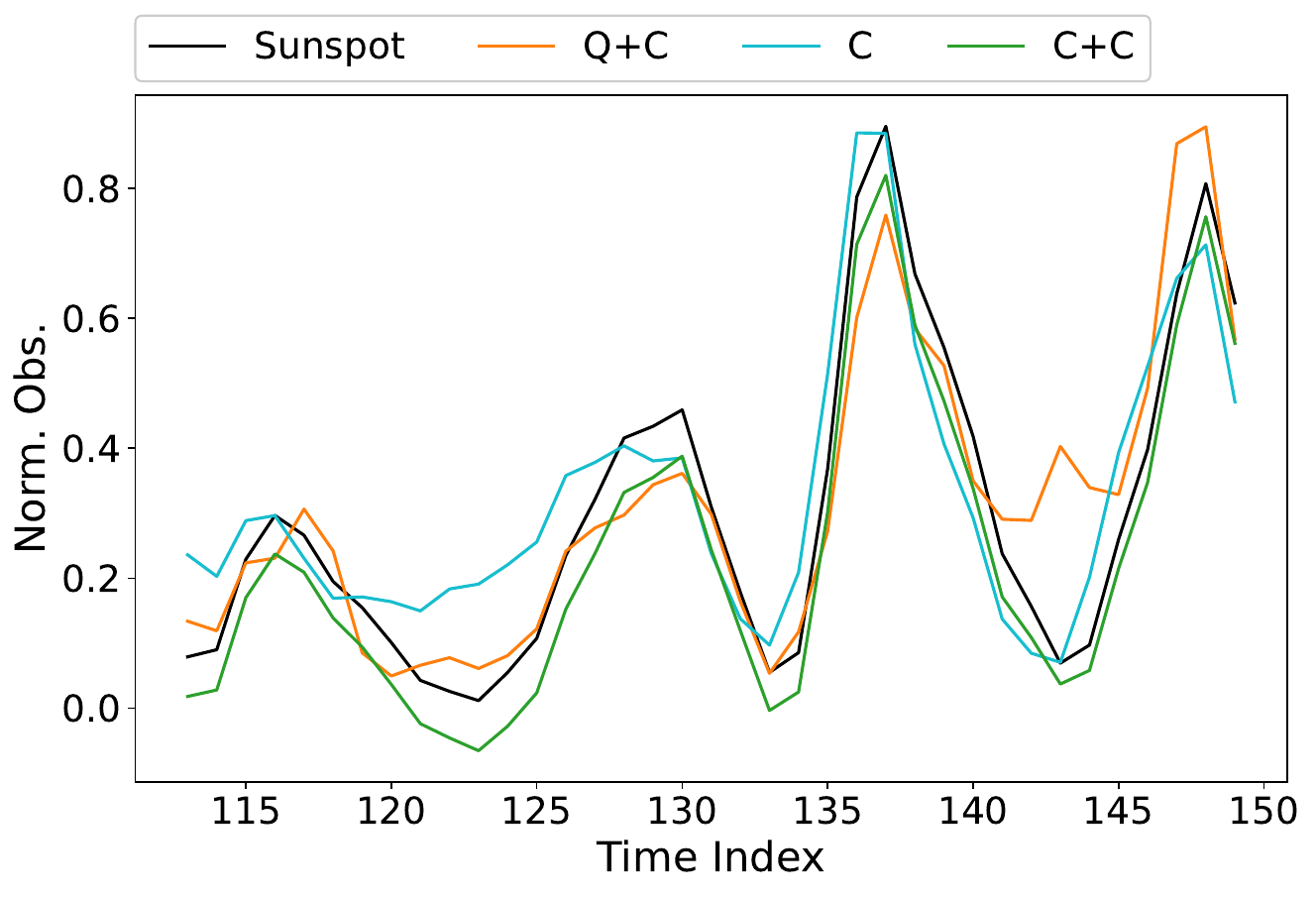}
        \caption{}
        \label{fig:sunspot-preds}
    \end{subfigure}
    \hspace{1cm}
    \begin{subfigure}{0.39\textwidth}
        \includegraphics[width=\textwidth]{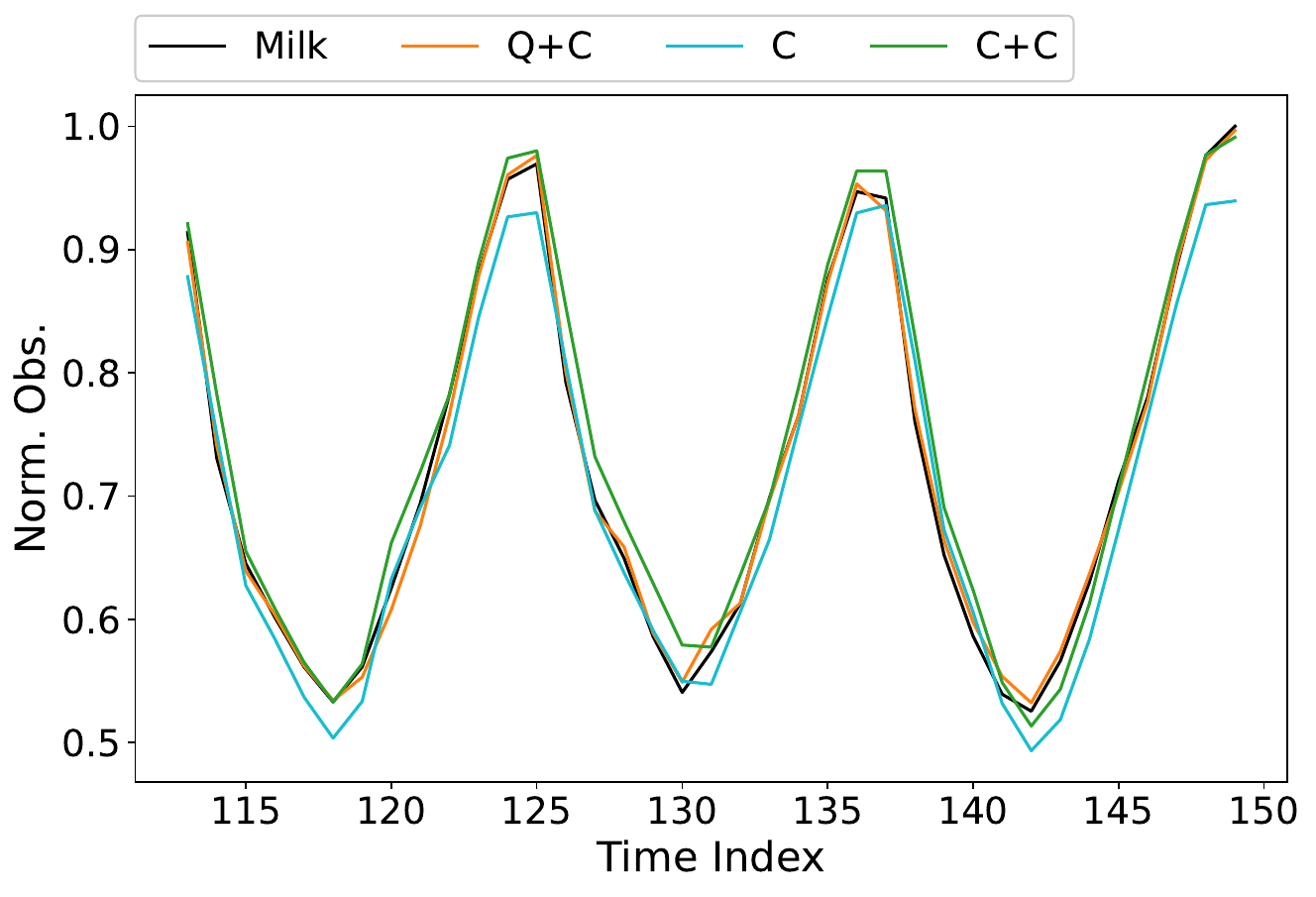}
        \caption{}
        \label{fig:milk-preds}
    \end{subfigure}
    
    \begin{subfigure}{0.39\textwidth}
        \includegraphics[width=\textwidth]{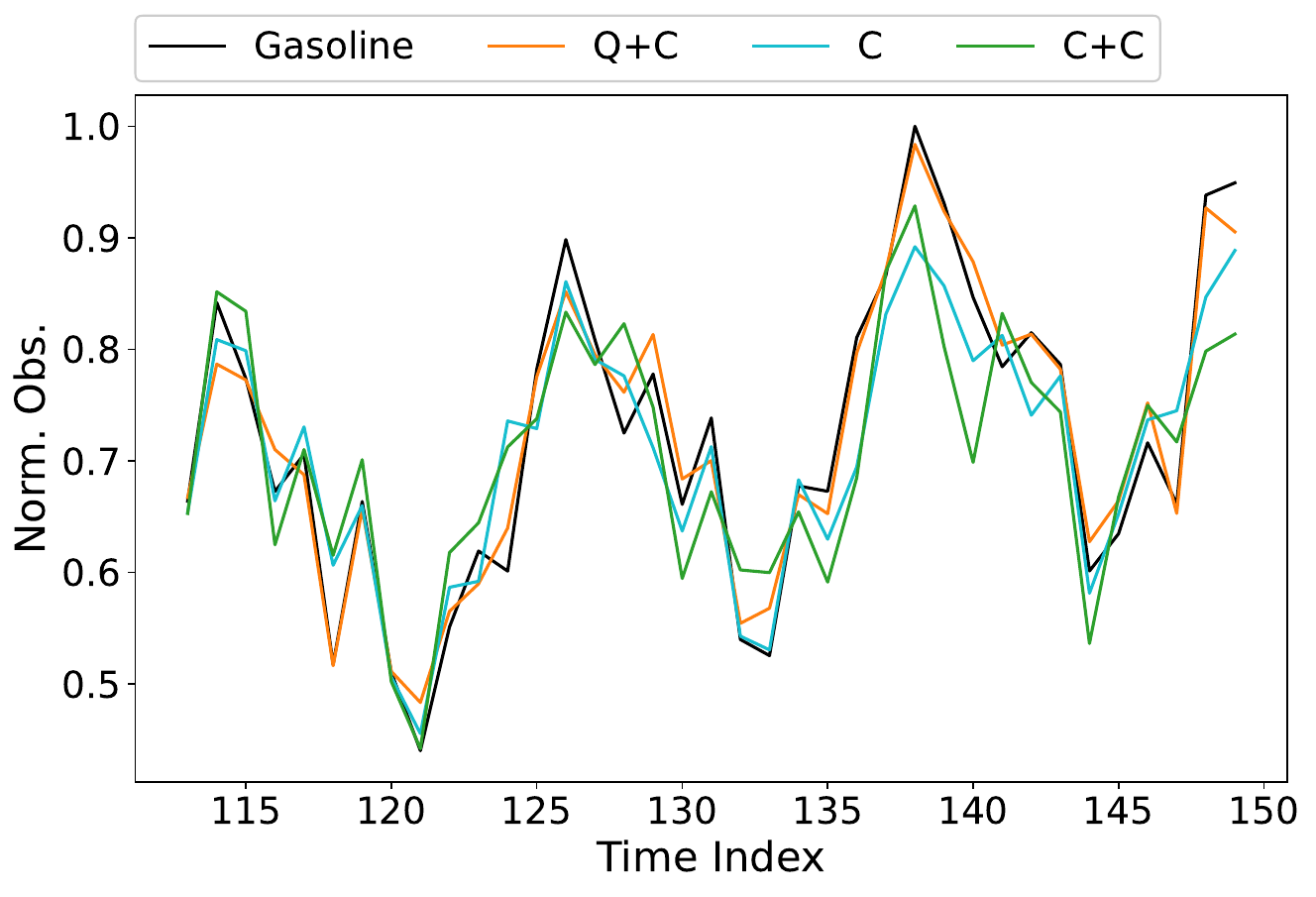}
        \caption{}
        \label{fig:gasoline-preds}
    \end{subfigure}
    \hspace{1cm}
    \begin{subfigure}{0.39\textwidth}
        \includegraphics[width=\textwidth]{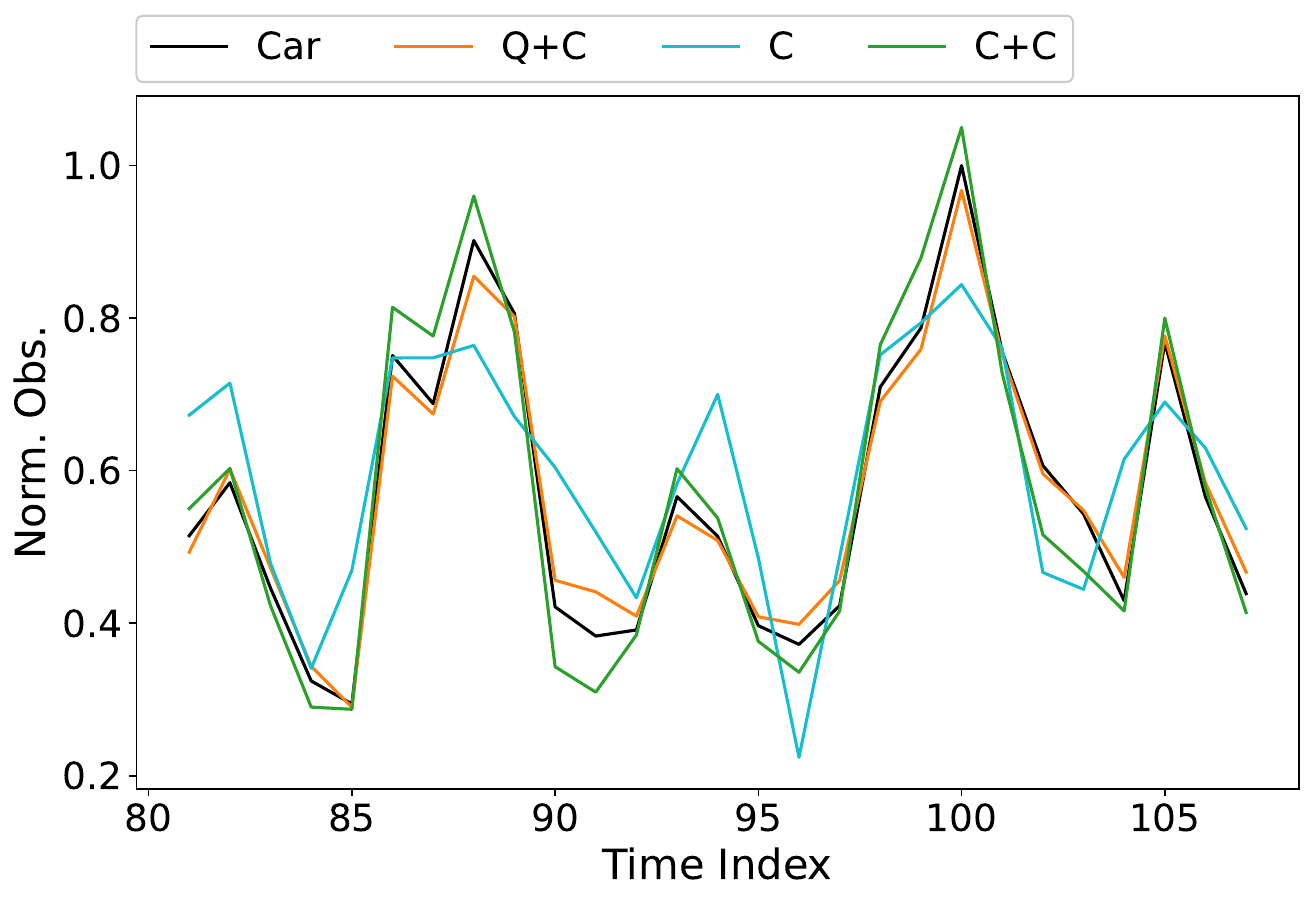}
        \caption{}
        \label{fig:car-preds}
    \end{subfigure}
    
    \begin{subfigure}{0.39\textwidth}
        \includegraphics[width=\textwidth]{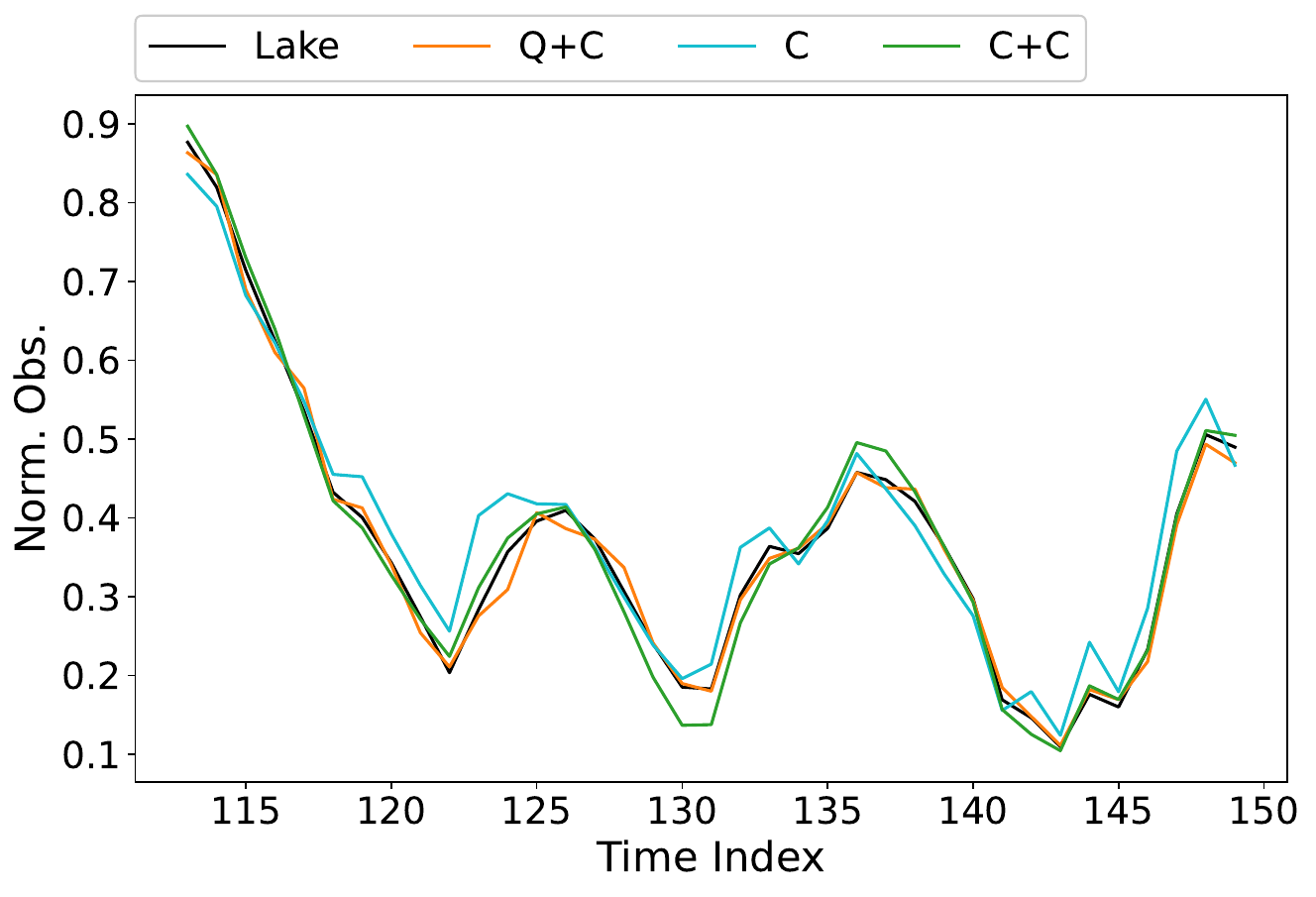}
        \caption{}
        \label{fig:lake-preds}
    \end{subfigure}

    \caption{Predictions of quantum-classical models and classical baselines in the seven problems. The target series and the three prediction series are represented by black, orange, cyan, and green lines. \textbf{a,b,c,d,e,f,g,} Predictions for the Temperature (\textbf{a}), Star (\textbf{b}), Sunspot (\textbf{c}), Milk (\textbf{d}), Gasoline (\textbf{e}), Car (\textbf{f}), and Lake (\textbf{g}) datasets.}
    \label{fig:preds}
\end{figure}

The quality of these predictions can be determined using metrics, such as MSE. To allow solid conclusions, the Mean Absolute Error~(MAE) and the Mean Absolute Percentage Error~(MAPE) are also used here. The supplementary material presents a discussion of error metrics. Table~\ref{tab:err-comp} compares Q+C with C and C+C in terms of these three metrics in each of the forecasting problems. For each pair of a problem and a metric, the best result between Q+C and C is underlined. When comparing with C+C, the best result is highlighted in bold. Note that MAPE values are multiplied by 100.

\begin{table}[ht]
    \caption{Error measures of quantum-classical models and classical baselines in the seven forecasting problems.}
    \label{tab:err-comp}
    \begin{tabularx}{\textwidth}{c|CCC|CCC|CCC}
        \hline
        & \multicolumn{3}{c|}{MSE} & \multicolumn{3}{c|}{MAE} & \multicolumn{3}{c}{MAPE} \\
        Dataset & Q+C & C & C+C & Q+C & C & C+C & Q+C & C & C+C \\
        \hline
        Temp. & \underline{0.0016} & 0.0032 & \textbf{0.0003} & \underline{0.0332} & 0.0478 & \textbf{0.0155} & \underline{6.90} & 15.14 & \textbf{3.26} \\
        Star & 0.0004 & \underline{0.0003} & \textbf{0.0001} & 0.0156 & \underline{0.0142} & \textbf{0.0071} & 6.59 & \underline{6.41} & \textbf{2.55} \\
        Sunspot & 0.0109 & \underline{0.0105} & \textbf{0.0044} & \underline{0.0768} & 0.0885 & \textbf{0.0650} & \textbf{\underline{58.95}} & 110.82 & 62.01 \\
        Milk & \textbf{\underline{0.0001}} & 0.0008 & 0.0007 & \textbf{\underline{0.0073}} & 0.0248 & 0.0201 & \textbf{\underline{1.07}} & 3.46 & 2.91 \\
        Gas. & \textbf{\underline{0.0007}} & 0.0029 & 0.0051 & \textbf{\underline{0.0213}} & 0.0416 & 0.0592 & \textbf{\underline{3.08}} & 5.65 & 8.18 \\
        Car & \textbf{\underline{0.0006}} & 0.0128 & 0.0024 & \textbf{\underline{0.0214}} & 0.0954 & 0.0411 & \textbf{\underline{4.21}} & 19.29 & 7.35 \\
        Lake & \textbf{\underline{0.0002}} & 0.0016 & 0.0005 & \textbf{\underline{0.0119}} & 0.0317 & 0.0172 & \textbf{\underline{3.39}} & 10.71 & 5.98 \\
        \hline
    \end{tabularx}
    \footnotetext{The MSE, MAE, and MAPE values obtained by Q+C in seven time series datasets are contrasted with the ones obtained by C and C+C. The best value between Q+C and C in respect of each metric in each dataset is underlined, while the best values between Q+C and C+C are highlighted in bold.}
\end{table}

The predictions basically followed each other in the Temperature problem, with differences in the peaks and valleys, as shown in Fig.~\ref{fig:temperature-preds}. Specifically in the valleys, Q+C was consistently better than C, which led to a gain of 50\% in terms of MSE, 31\% in terms of MAE, and 54\% in terms of MAPE, considering the performance of Q+C against C in this problem. However, C+C was better than Q+C according to all metrics in Temperature, especially due to the predictions for the highest and lowest observations of the target time series. In the Star case, although the metrics indicate that Q+C was worse than both C and C+C, Fig.~\ref{fig:star-preds} shows that all predictions essentially captured the dynamics of the problem.

On the other hand, the predictions for Sunspot shown in Fig.~\ref{fig:sunspot-preds} differ from each other and from the target series to some extent. In this case, Q+C and C alternated between providing the best prediction throughout the target time series, leading to similar values of MSE and MAE, although these metrics disagree on the best overall model between Q+C and C. As a result of the MAE tolerance to large errors, Q+C was better than C in terms of MAE, with a gain of 13\% in performance. In turn, especially because Q+C better predicted the minimum values of the target time series, a performance gain of 47\% is even obtained when comparing Q+C to C in view of the MAPE metric in this problem. Furthermore, a better approximation of the lowest points also caused Q+C to emerge as better than C+C in terms of MAPE, even though the predictions of C+C better follow the overall dynamics of the Sunspot problem. These metric divergences in Sunspot are further discussed in the supplementary material.

Fig.~\ref{fig:milk-preds} presents the predictions for the Milk problem, where Q+C accurately followed the target series, principally in the peaks and valleys. In this problem, Q+C represents a performance gain of 88\% in terms of MSE, 71\% in terms of MAE, and 69\% in terms of MAPE when compared to C. Compared to C+C, the gains are 86\%, 64\%, and 63\%, respectively. The Gasoline problem shown in Fig.~\ref{fig:gasoline-preds} illustrates a case where the models alternated between providing the best prediction from a point to another, although Q+C provided the best predictions in mean terms. In Gasoline, the performance gains of Q+C over C are 76\%, 49\%, and 45\% based on each respective metric. The gains of Q+C against C+C in Gasoline are 86\%, 64\%, and 62\%.

Finally, Figs.~\ref{fig:car-preds} and~\ref{fig:lake-preds} show for Car and Lake that, while C struggled to capture the dynamics of the problems, Q+C and C+C followed the two target time series, with Q+C approximating the true values even further in general. Compared to C in Car, Q+C represents performance gains of 95\% in terms of MSE and 78\% in terms of both MAE and MAPE. Against C+C, the gains are 75\%, 48\% and 43\% in Car. For the Lake problem, the gains over C and C+C are, respectively, 88\% and 60\% in MSE, 62\% and 31\% in MAE, and 68\% and 43\% in MAPE.
\section{Discussion}

In summary, compared to classical single models in seven forecasting problems, the proposed integration of quantum and classical models through the error correction method provided better results in five problems in terms of MSE and six problems in terms of MAE and MAPE. In these cases, the mean performance gains were 79\%~$\pm$~18\% in terms of MSE, 51\%~$\pm$~25\% in terms of MAE, and 60\%~$\pm$~13\% in terms of MAPE. Even when two classical models are combined in an analogous way, the proposed quantum-classical integration still provided the best results in four cases based on the MSE and MAE metrics and in five cases based on MAPE. On average, the performance gains in these cases were 77\%~$\pm$~12\%, 52\%~$\pm$~16\%, and 43\%~$\pm$~23\% according to the three metrics, respectively.

Such evidence reveals that optimal combinations of quantum and classical models under the error correction method are able to handle time patterns that go beyond the capacity of classical baselines, even if the classical baselines are also optimized for the problem. Both classical single models and error-correction-based classical-classical models were unable to surpass the performance of error-correction-based quantum-classical models in the majority of cases according to three metrics. Primary patterns captured by quantum models from the original time series, in conjunction with additional patterns captured by classical models from the quantum errors, can be used to produce considerably better predictions. The proposed hybrid models took advantage of an established hybridization scheme to leverage quantum phenomena to capture complementary data relationships in forecasting tasks. As experimentally demonstrated that the classical capacity can be augmented in the proposed way, this work opens a promising path supported by the idea of inserting quantum models within the extensive error correction framework for time series forecasting.

Possible directions to build on the advances already provided here include using other operators such as weighted sum and even nonlinear functions to explore further relations between the quantum and classical predictions. Arrangements that consider quantum models, classical models, and also statistical models, in different orders, should be investigated in pursuit of a still broader capacity to handle time patterns. More quantum models can be applied in more forecasting problems, especially scaling these models and exploring strategies to circumvent plateaus in the optimization landscapes. Enlightening the actual potential of these models in time series forecasting can generate data to study the class of patterns that quantum models are likely to capture well. This insight can guide the use of quantum models together with other paradigms of models in forecasting tasks, which can surpass the error correction method, achieving alternatives such as static and dynamic ensembles of models.
\section{Methods}

\subsection{Quantum models}

\begin{figure}[ht]
    \centering
    \begin{quantikz}[thin lines]
        \lstick{\ket{0}} \slice[style=blue]{1} & \gate[3, style={fill=blue!20}]{E \big( x_0, \cdots, x_{n_Q - 1} \big) } & \linethrough \midstick[3, brackets=left]{$\times R$} \slice[style=blue]{2} & \gate[3, style={fill=blue!20}]{G \big( \theta_0^s, \cdots, \theta_{p-1}^s \big) } & \linethrough \midstick[3, brackets=left]{$\times S$} \slice[style=blue]{3} & \meter[3, style={fill=gray!20}]{} \\ [1cm]
        \setwiretype{n} \lstick{\push{\raisebox{1.5ex}{\ \ \vdots\ \ }}} & & & & & \\ [1cm]
        \lstick{\ket{0}} & & \linethrough & & \linethrough &
    \end{quantikz}
    \caption{General quantum circuit for machine learning tasks. The circuit basically contains two inner parts that are delimited with the aid of vertical dashed lines. The first part, represented from slice 1 to slice 2, is responsible for encoding the classical input data $x_0, \cdots, x_{n_Q-1}$ via an operator $E(\cdot)$ that can be repeated $R$ times. The second part, from slice 2 to slice 3, acts on the quantum-encoded input data via an operator $G(\cdot)$ that is composed of the parameterized gates to be adjusted iteratively. Each repetition of the operator $G(\cdot)$ is dependent on a different set of parameters $\theta_0^s, \cdots, \theta_{p-1}^s$, where $s = 0, \cdots, S-1$. Such a repeated application of the operator $G(\cdot)$ results in a final quantum state from which measurements give the circuit output.}
    \label{fig:general-vqc}
\end{figure}
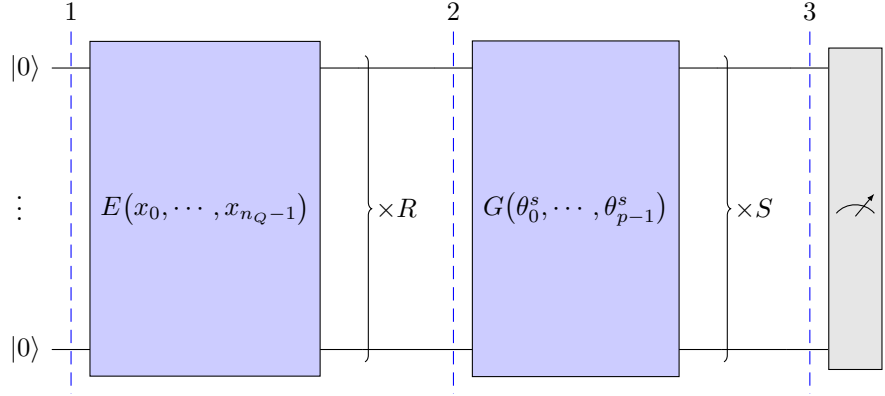

Currently, the most promising quantum framework for machine learning is based on the idea of producing the expected outputs of a task with parameterized quantum circuits that can be adjusted in an iterative manner~\cite{mitarai_1st-vqc}. A quantum circuit can then be subdivided into a first inner circuit that encodes the input data and a second inner circuit that actually contains the parameterized gates to be adjusted. Fig.~\ref{fig:general-vqc} shows a general quantum circuit under this learnable scheme, where vertical dashed lines delimit the inner circuits. Data encoding is represented from slice 1 to slice 2, while the learnable part covers from slice 2 to slice 3.

The circuit starts encoding the classical input data $x_0, \cdots, x_{n_Q - 1}$ into the quantum space via an operator $E(\cdot)$ that can be repeated $R$ times. Subsequently, $G(\cdot)$ operates on the quantum-encoded input data using a set of gates that depend on different parameters $\theta_0^s, \cdots, \theta_{p-1}^s$ in each learnable layer $s$, where $s = 0, \cdots, S-1$. In the experiments conducted here, three pairs $E(\cdot)\text{-}G(\cdot)$ were explored, which can be found in ref.~\cite{chen_vqc-lstm-ops}, ref.~\cite{ceschini_lstm-fc-vqc}, and ref.~\cite{ruiz_fc-vqc-fc}, respectively. The first of these pairs, named here $E_{YZ}G_U$, is defined as the composition of the operators:
\begin{align*}
    E_{YZ} &= \bigotimes_{q=0}^{n_Q-1} \bigg(
                    R_z \Big( \text{arctan} \big( x_q^2 \big) \Big) \cdot
                    R_y \Big( \text{arctan} \big( x_q \big) \Big) \cdot
                    H
                \bigg) \text{ and} \\[10pt]
    G_U &=
            \bigotimes_{q=0}^{n_Q-1} U \big( \theta_{3q}^s, \theta_{3q+1}^s, \theta_{3q+2}^s \big) \cdot
            \prod_{q=0}^{n_Q-1} CX_{q,q+2} \cdot
            \prod_{q=0}^{n_Q-1} CX_{q,q+1},
\end{align*}

\noindent
where $CX_{i,j}$ controls the qubit $i$ to apply the $X$ gate on the qubit $j \bmod n_Q$, while preserving the other qubits. The second pair $E(\cdot)\text{-}G(\cdot)$ used in the pool of quantum models, called $E_{PP}G_{YZ}$ here, is formed by the operators:
\begin{align*}
    E_{PP} &=
                \prod_{q=0}^{n_Q-2} P^*_{q,q+1} \big( x_q, x_{q+1} \big) \cdot
                \bigotimes_{q=0}^{n_Q-1} \Big( P \big( 2x_q \big) \cdot H \Big)
            \text{ and} \\[10pt]
    G_{YZ} &= \Bigg(
                    \bigotimes_{q=0}^{n_Q-1} R_z \big( \theta_{3n_Q+q}^s \big) \cdot
                    \bigotimes_{q=0}^{n_Q-1} R_y \big( \theta_{2n_Q+q}^s \big) \cdot
                    \prod_{q=0}^{n_Q-2} CX_{q,q+1} \\
                    &\qquad \cdot
                    \bigotimes_{q=0}^{n_Q-1} R_z \big( \theta_{n_Q+q}^s \big) \cdot
                    \bigotimes_{q=0}^{n_Q-1} R_y \big( \theta_q^s \big)
                \Bigg),
\end{align*}

\noindent
where $P^*_{i,j}(x_i, x_j)$ applies $CX_{i,j}$, then $P\big( 2(\pi - x_i)(\pi - x_j) \big)$ in the qubit $j$, and finally another $CX_{i,j}$, while preserving the other qubits. The third and final quantum model, $E_YG_{XYZ}$, originates from the following definitions:
\begin{align*}
    E_Y &=
                \bigotimes_{q=0}^{n_Q-1} R_y \big( \pi x_q \big)
            \text{ and} \\[10pt]
    G_{XYZ} &=
                \prod_{q=0}^{n_Q-2} CX_{q,q+1} \cdot
                \bigotimes_{q=0}^{n_Q-1} R_z \big( \theta_{2n_Q+q}^s \big) \cdot
                \bigotimes_{q=0}^{n_Q-1} R_y \big( \theta_{n_Q+q}^s \big) \cdot
                \bigotimes_{q=0}^{n_Q-1} R_x \big( \theta_q^s \big).
\end{align*}

After encoding a classical input data through a particular $E(\cdot) \times R$ and then transforming the quantum-encoded input data via a particular $G(\cdot) \times S$, the result is a final quantum state $\ket{\psi_f}$. A circuit output is obtained by computing the expectation value of an observable, which actually weighs the eigenvalues of such an observable by the probabilities of $\ket{\psi_f}$ being measured in the corresponding basis states. Here, the observable was the operator $Z^{\otimes n_Q}$, and thus $\bra{\psi_f} Z^{\otimes n_Q} \ket{\psi_f}$ produced the circuit output as a weighted sum of the eigenvalues +1 and -1 considering the probabilities of $\ket{\psi_f}$ in the computational basis.

Since the quantum outputs are between -1 and 1, the points of each forecasting problem were scaled to this interval of values. However, primary experiments revealed an inability of quantum models to predict values beyond the subinterval of the training data points, even though the values to be predicted are in the interval [-1, 1]. This issue clearly occurred when predicting time series with a trend. To circumvent such a limitation, each input window was shifted so that the windows became centered in the same midpoint~\cite{emmanoulopoulos_vqc-trend-tsf}. A centralization around zero was obtained by subtracting the median value of each input window from the values of the input window. Each expected output was also subtracted by the median value of the respective input window. Let $m_t^{n_Q}$ be the median value of the input window $y_{ \{ t - (n_Q - 1) \} }, \cdots, y_t$. That is, $m_t^{n_Q}$ is the median value of the $n_Q$-sized window ending in $y_t$. Thus, the window that is actually given as input to the quantum model is $y^*_{ \{ t - (n_Q - 1) \} }, \cdots, y^*_t$, where $y^*_i = y_i - m_t^{n_Q}$. Consequently, the expected output for this input window is $y^*_{t+1} = y_{t+1} - m_t^{n_Q}$.

In this way, the objective of the optimization process becomes to find the parameters that make the quantum model approximate the shifted expected outputs from the shifted input windows. Five instances of each quantum model were trained in each forecasting problem. Each model instance is defined by a hyperparameter configuration that was randomly sampled from the space delimited in Table~\ref{tab:q-hyperparam}, where the optimizer $O$, with the learning rate $\eta$, minimized the MSE metric in the training set during 1000 epochs. An early stopping was triggered in the training process whenever the MSE value obtained in the validation set increased for 10 consecutive epochs. These quantum models were implemented using Qiskit Machine Learning~\cite{sahin_qiskit-ml} integrated with PyTorch~\cite{paszke_pytorch}. This integration finally allowed the use of PyTorch-Ignite to train and evaluate the models~\cite{fomin_pytorch-ignite}.

\begin{table}[ht]
    \caption{Space from which the hyperparameter configurations of the quantum models were randomly sampled.}
    \label{tab:q-hyperparam}
    \begin{tabularx}{\textwidth}{C|C}
        \hline
        Hyperparameter & Values \\
        \hline
        $n_Q$ & [3, 4, 5, 6] \\
        $R$ & [1, 2, 3] \\
        $S$ & [1, 2, 3, 4, 5, 6, 7, 8] \\
        $O$ & [SGD, Adam] \\
        $\eta$ & [0.01, 0.05, 0.1, 0.25, 0.5] \\
        \hline
    \end{tabularx}
    \footnotetext{Each configuration is defined by a number $n_Q$ of inputs, a number $R$ of encoding layers, a number $S$ of learnable layers, an optimizer $O$, and a learning rate $\eta$.}
\end{table}

The model instance with the best validation MSE after the training process was chosen to represent the quantum model in the given problem. Then, the respective median values were properly added to the predictions of such a best instance to obtain the quantum predictions for the original dynamics of the target time series, even if a trend exists. Finally, the target time series was rescaled from the interval [-1, 1] to the interval [0, 1], and the quantum predictions were transformed accordingly. Here, classical models learned to correct the best instance of each quantum model only in the interval [0, 1], which is also the interval where the results were presented.

\subsection{Classical models}

The pool of classical models was composed of three well-established artificial neural networks. The first of these models was the MLP model, a feedforward network formed by three layers of neurons: an input layer, a hidden layer, and an output layer~\cite{bishop_ml-book}. The number of inputs for the MLP model in a given problem, which determines the number of input neurons, was defined as (1) the highest significant lag in the first 20 lags of the partial autocorrelation function or (2) 20 inputs if there were no significant lags. The output layer contained only one neuron. Regarding the hidden layer, the number of neurons and the activation functions were hyperparameters to be defined.

The second evaluated model was the LSTM model, a recurrent neural network with gating mechanisms that control the flow of information in long- and short-term memory states~\cite{hochreiter_lstm}. The state size and the number of layers for the LSTM model were optimized as hyperparameters. Finally, the third model contained in the pool was the NBEATS model, a deep neural network composed of multiple stacks of MLP models that are connected through residual backward links and aggregating forward links~\cite{oreshkin_nbeats}. The adopted configuration for the NBEATS model stacked three MLP models, each with two hidden layers, where the number of neurons for both hidden layers was optimized as a hyperparameter. For both LSTM and NBEATS, the number of inputs was defined based on the partial autocorrelation function, as in the MLP model.

Table~\ref{tab:c-hyperparam} shows the space of values for the hyperparameters of each classical model. Given a problem, 10 hyperparameter configurations were experimented for each model, where the optimizer was a hyperparameter in the MLP model only. The spaces of learning rates are actually continuous intervals that were still mapped to the logarithmic domain to better sample values between different orders of magnitude. Each instance of an MLP model was trained for 1000 epochs, while LSTM and NBEATS instances were trained for 500 epochs. A training process stopped early if there were no validation improvements during 10 epochs for MLP instances and 25 epochs for LSTM and NBEATS instances. In terms of implementation, Scikit-learn~\cite{pedregosa_scikit-learn} was used for the MLP model, whereas NeuralForecast~\cite{olivares_neuralforecast} was used for the LSTM and NBEATS models. The hyperparameters of all three models were optimized using Optuna~\cite{akiba_optuna}.

\begin{table}[ht]
    \caption{Spaces from which the hyperparameter values of each classical model were sampled.}
    \label{tab:c-hyperparam}
    \begin{tabularx}{\textwidth}{c|C|C}
        \hline
        Model & Hyperparameter & Values \\
        \hline
        \multirow{4}{*}{MLP} & No. hidden neurons & [2, 5, 10, 15, 20, 50] \\
        & Activation function & [Logistic, ReLU] \\
        & Optimizer & [Adam, L-BFGS] \\
        & Learning rate & [$10^{-4}$, $10^{-1}$] \\
        \hline
        \multirow{3}{*}{LSTM} & State size & [64, 128, 256] \\
        & No. layers & [1, 2] \\
        & Learning rate & [$10^{-5}$, $10^{-1}$] \\
        \hline
        \multirow{2}{*}{NBEATS} & No. hidden neurons & [64, 128, 256, 512] \\
        & Learning rate & [$10^{-5}$, $10^{-1}$] \\
        \hline
    \end{tabularx}
\end{table}

For each model, the trained instance with the best MSE value in the validation set was chosen. In this way, only the predictions of such the best instance were considered for the given problem, regardless of whether the task is to predict the original time series or to predict forecasting errors. However, since the forecasting errors were originally around zero, the classical models were trained to predict errors considering the errors themselves being scaled to the interval [0, 1]. Thus, such a scaling needed to be undone before using the error predictions of the best model instance to make corrections in another model. Moreover, the MSE values in the validation set were finally used to define the C, Q+C, and C+C models as the best model among all the classical single models, all the quantum-classical hybrid models, and all the classical-classical hybrid models, respectively.

\subsection{Datasets}

In this study, seven well-known time series forecasting problems were considered. Table~\ref{tab:data} summarizes the differences between the problems, especially in terms of the application domain and sampling frequency. Together, these problems provided a broad testbed for assessing the capacity of the proposed models against classical baselines in distinct temporal dynamics. All datasets were obtained from the Time Series Data Library~\cite{hyndman_tsdl}.

\begin{table}[ht]
    \caption{Forecasting problems where the model capacities were assessed.}
    \label{tab:data}
    \begin{tabularx}{\textwidth}{c|c|C|c|c}
        \hline
        Dataset & Subject & Description & Frequency & Length \\
        \hline
        Temperature & Meteorology & Air temperature in Nottingham & Monthly & 240 \\
        Star & Physics & Brightness of a star on midnights & Daily & 600 \\
        Sunspot & Physics & Wolf’s number of sunspots & Annual & 289 \\
        Milk & Agriculture & Production of milk per cow & Monthly & 156 \\
        Gasoline & Sales & Demand of gasoline in Ontario & Monthly & 192 \\
        Car & Sales & Sale of cars in Quebec & Monthly & 108 \\
        Lake & Hydrology & Level of the lake Erie & Monthly & 600 \\
        \hline
    \end{tabularx}
\end{table}

\begin{figure}[htp!]
    \centering
    
    \begin{subfigure}{0.39\textwidth}
        \includegraphics[width=\textwidth]{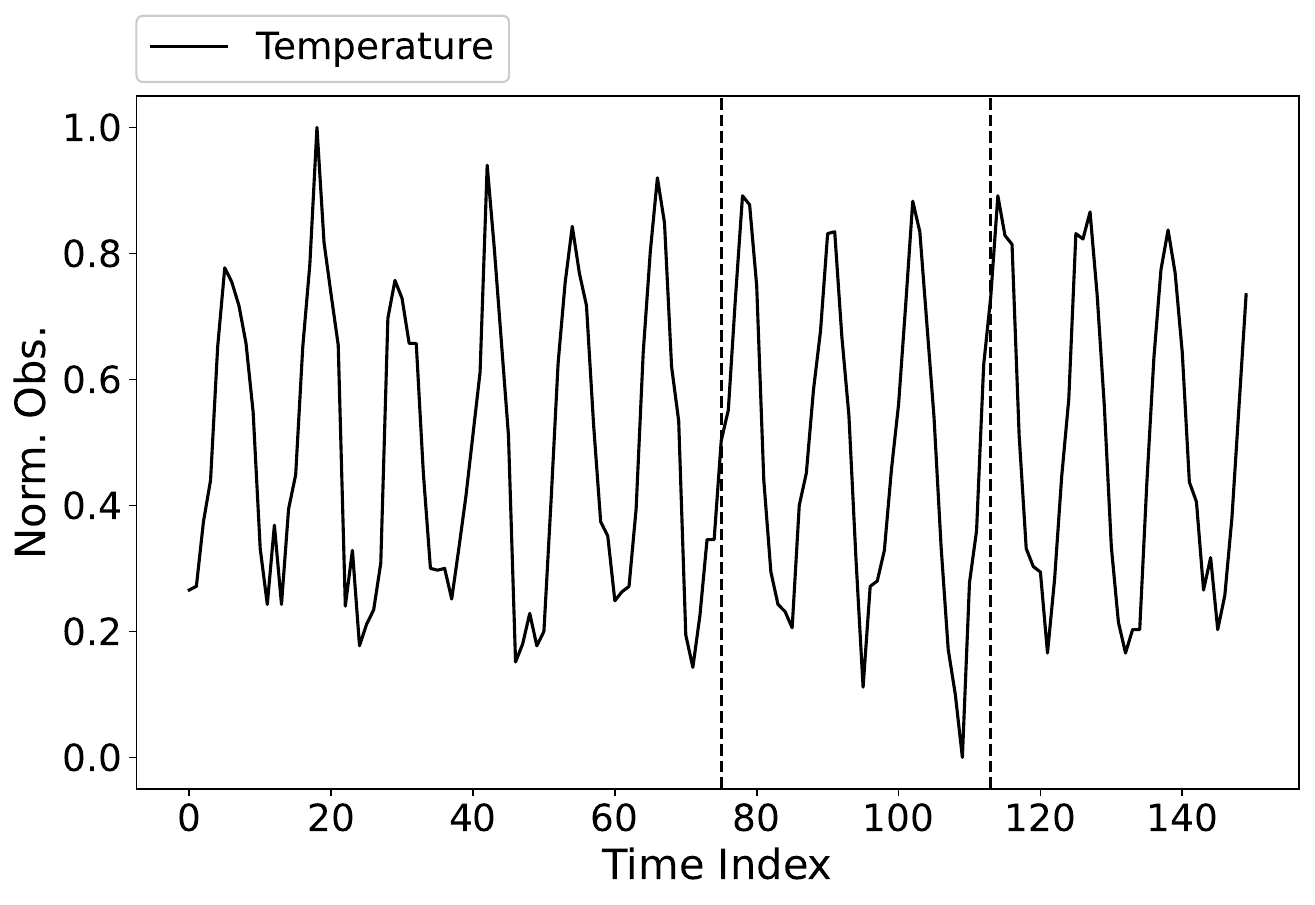}
        \caption{}
        \label{fig:temperature-prob}
    \end{subfigure}
    \hspace{1cm}
    \begin{subfigure}{0.39\textwidth}
        \includegraphics[width=\textwidth]{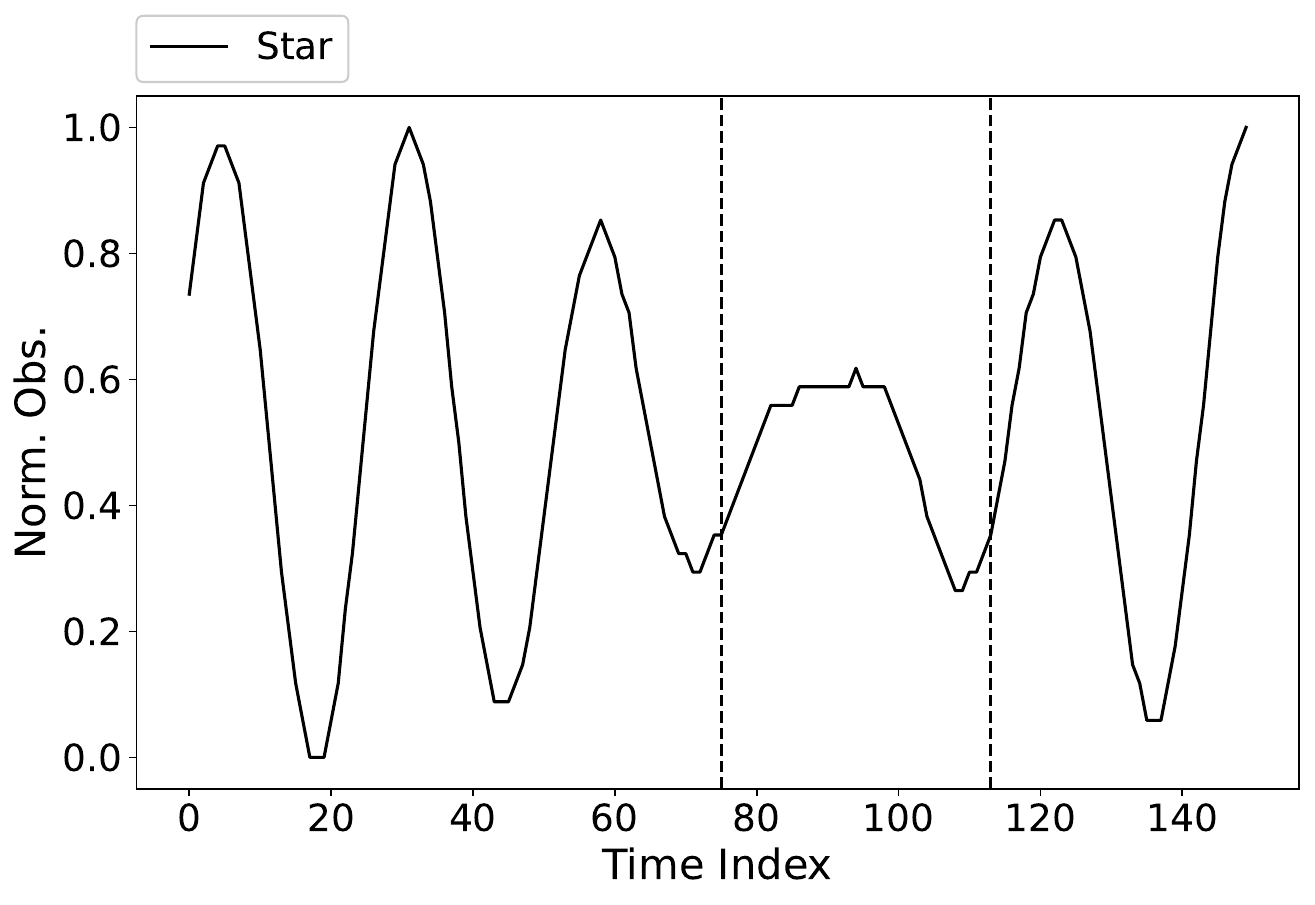}
        \caption{}
        \label{fig:star-prob}
    \end{subfigure}
    
    \begin{subfigure}{0.39\textwidth}
        \includegraphics[width=\textwidth]{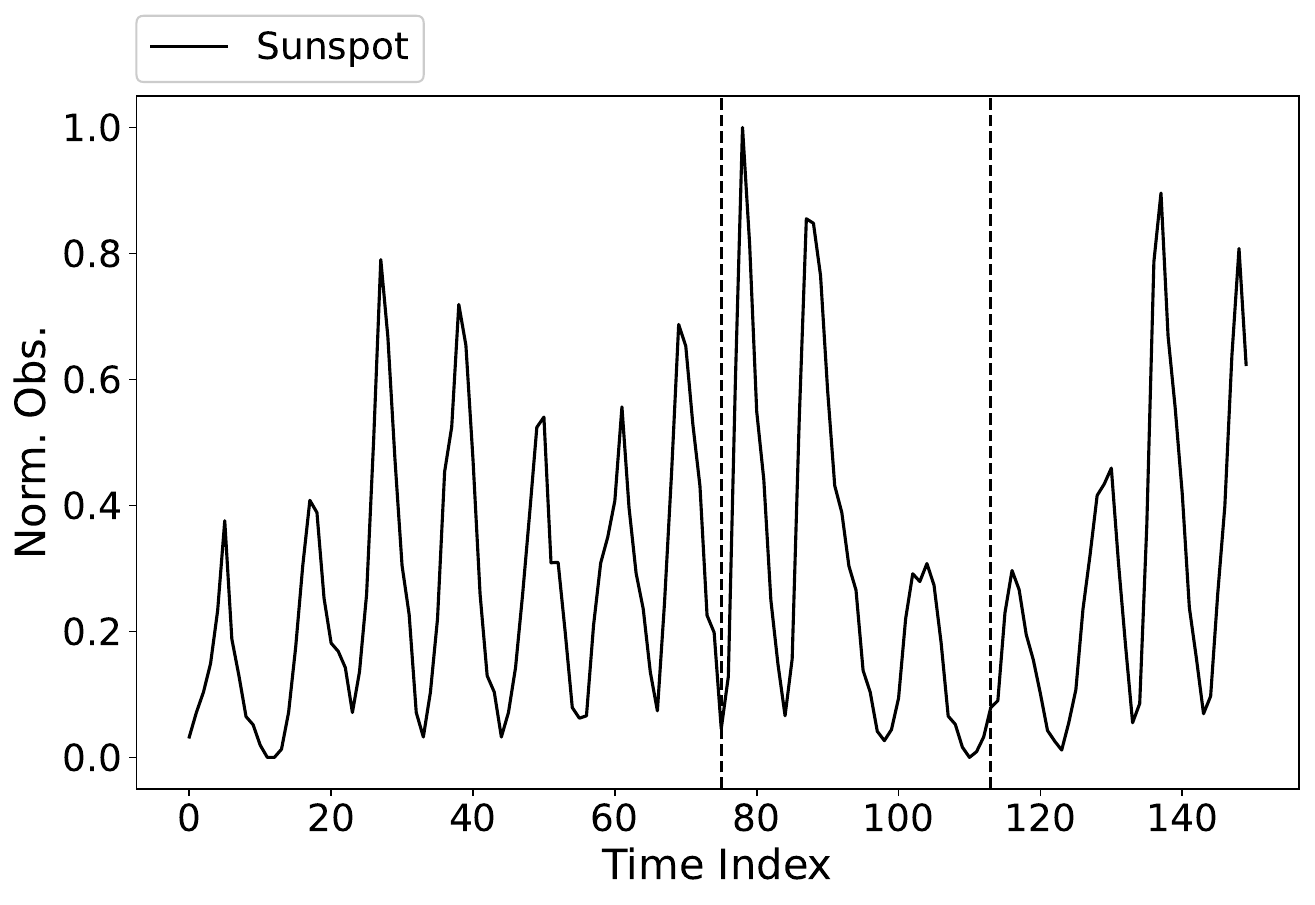}
        \caption{}
        \label{fig:sunspot-prob}
    \end{subfigure}
    \hspace{1cm}
    \begin{subfigure}{0.39\textwidth}
        \includegraphics[width=\textwidth]{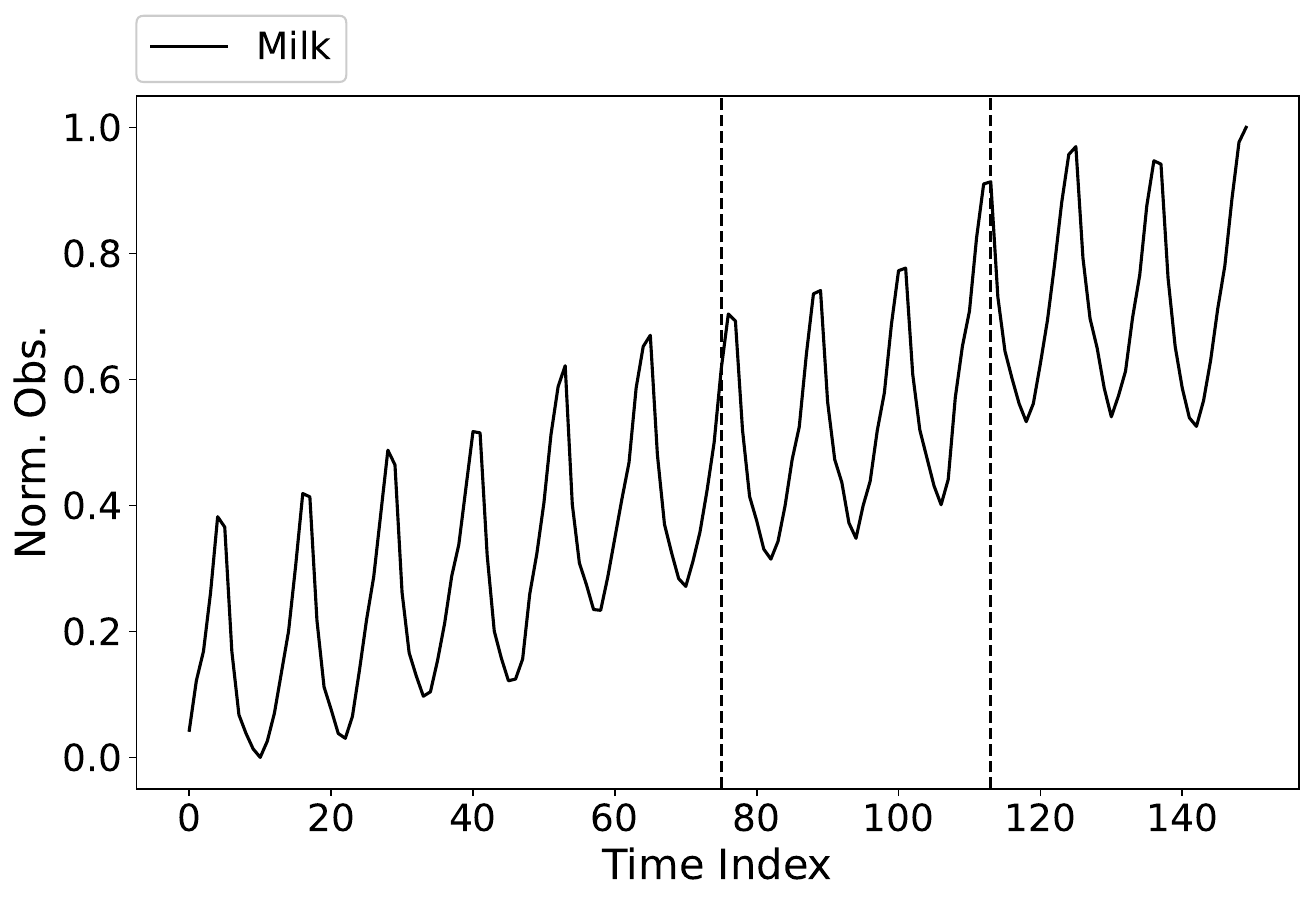}
        \caption{}
        \label{fig:milk-prob}
    \end{subfigure}
    
    \begin{subfigure}{0.39\textwidth}
        \includegraphics[width=\textwidth]{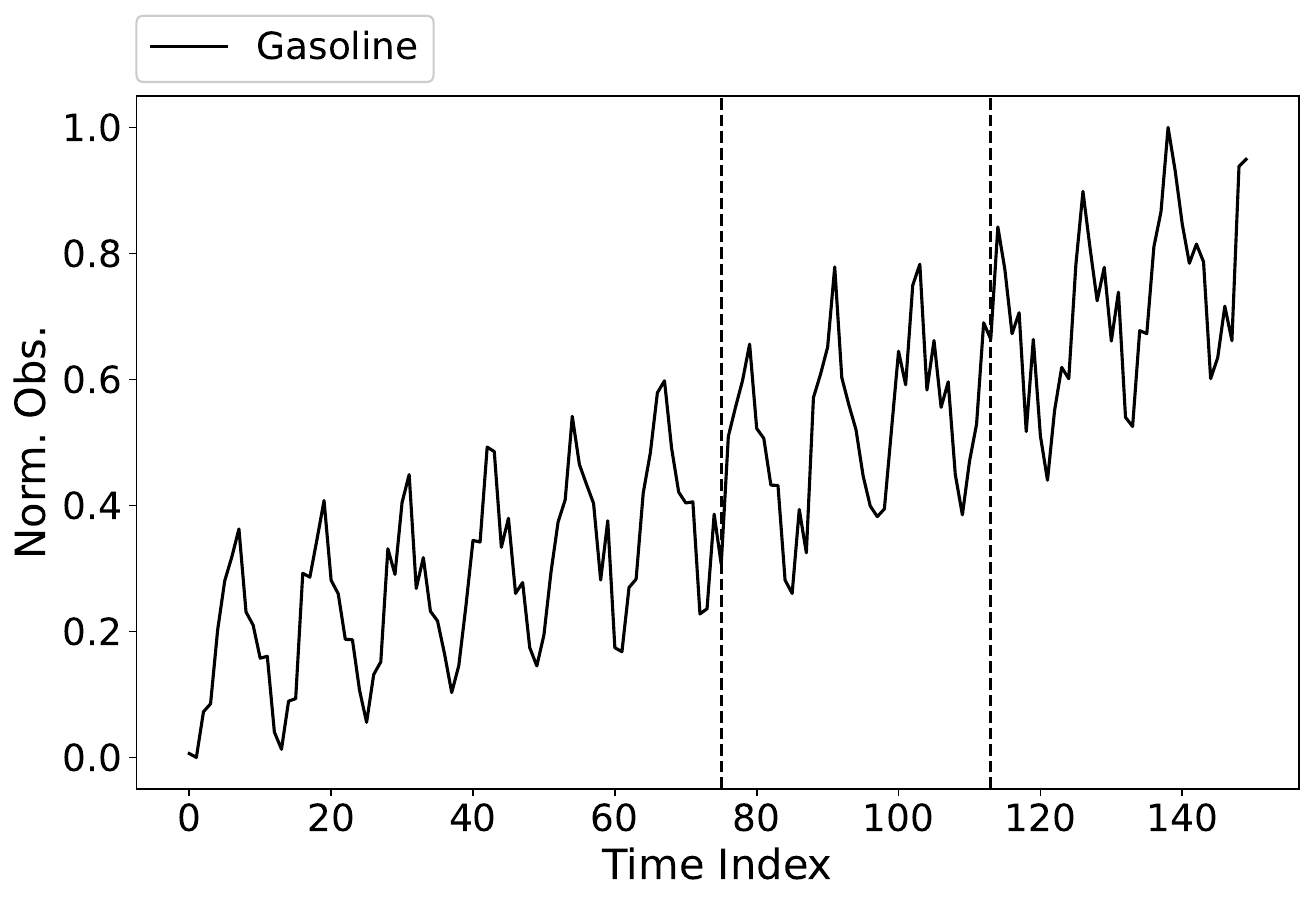}
        \caption{}
        \label{fig:gasoline-prob}
    \end{subfigure}
    \hspace{1cm}
    \begin{subfigure}{0.39\textwidth}
        \includegraphics[width=\textwidth]{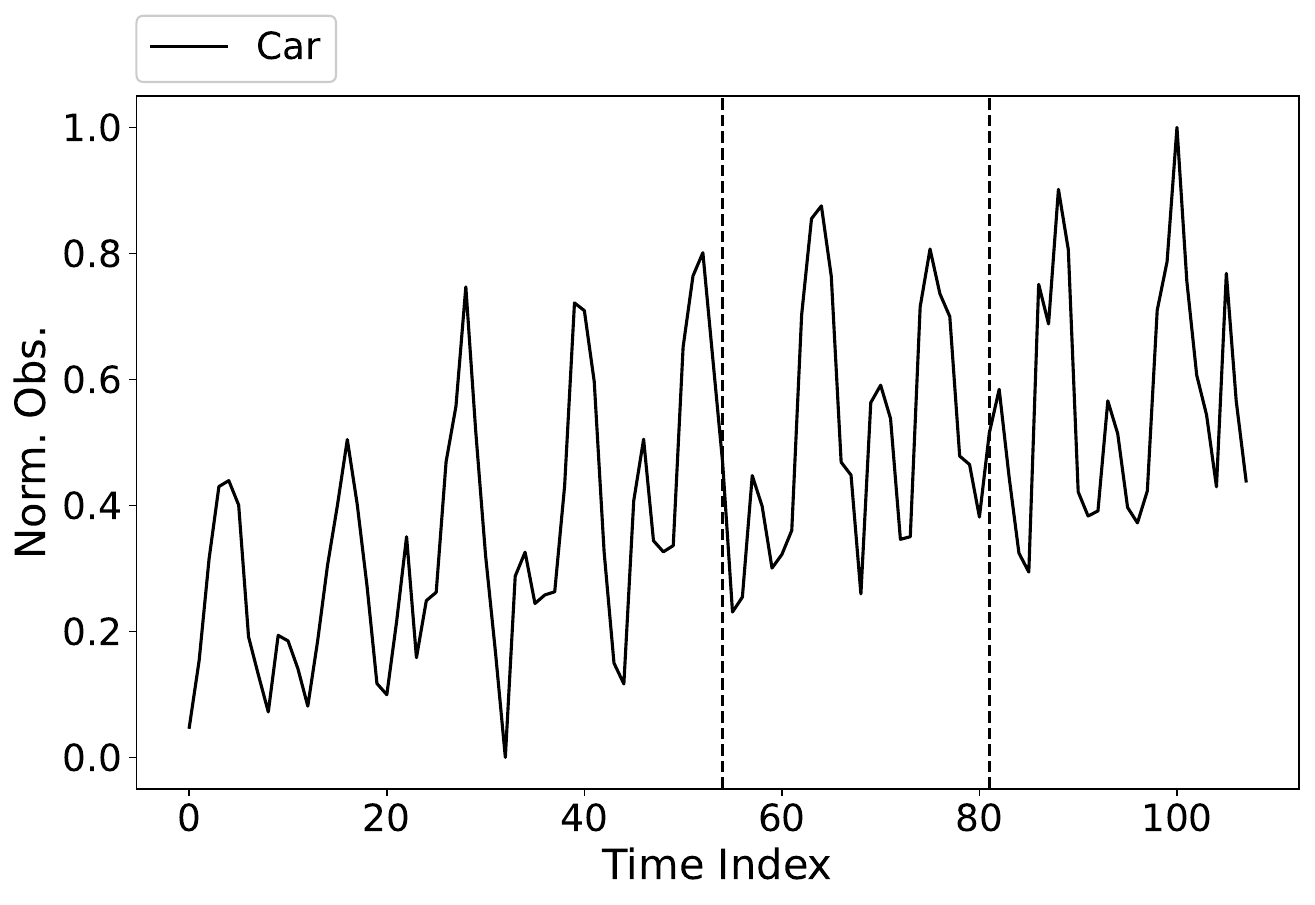}
        \caption{}
        \label{fig:car-prob}
    \end{subfigure}
    
    \begin{subfigure}{0.39\textwidth}
        \includegraphics[width=\textwidth]{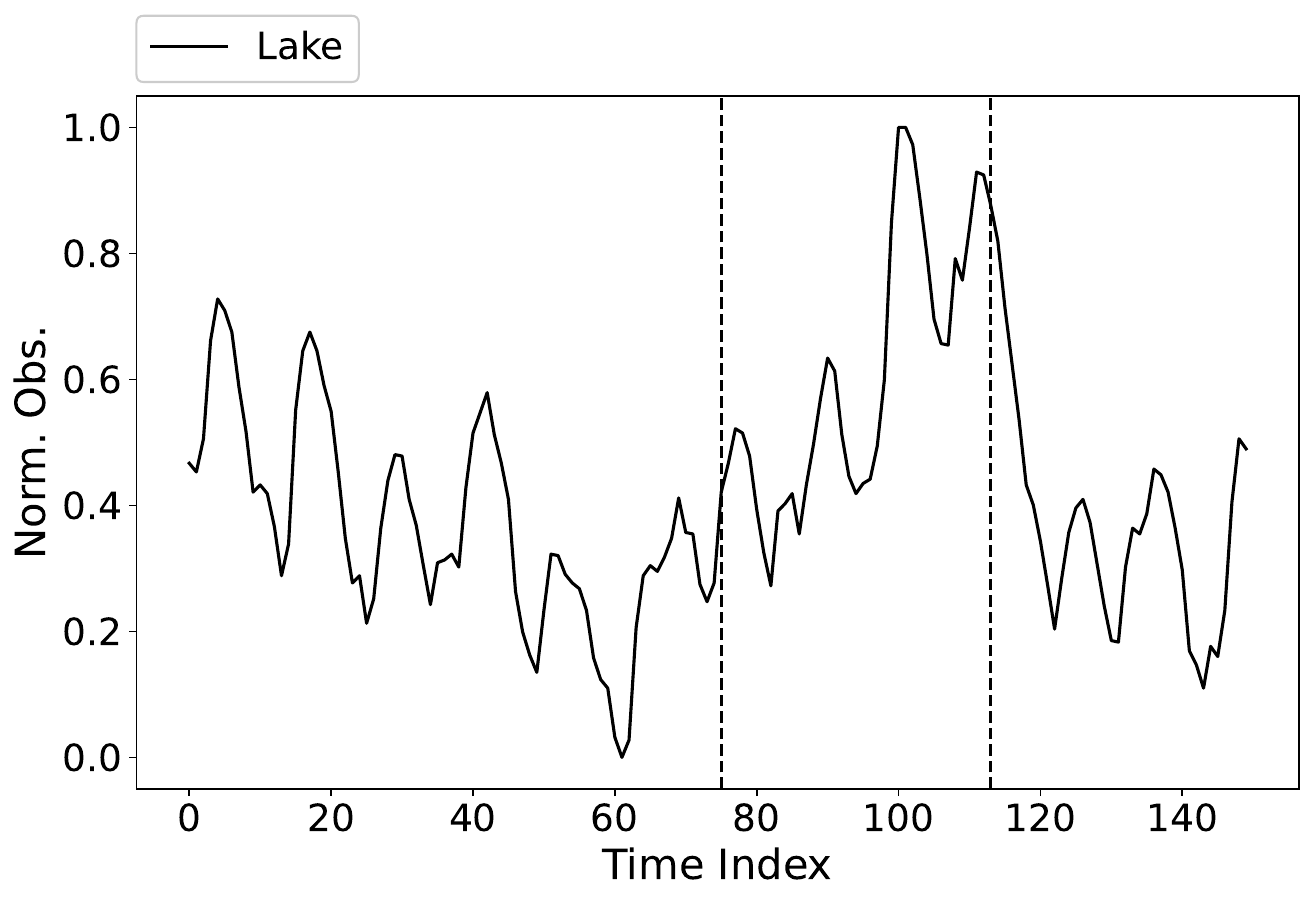}
        \caption{}
        \label{fig:lake-prob}
    \end{subfigure}

    \caption{Observations of the seven forecasting problems used to evaluate the models. The two vertical dashed lines separate the training, validation and test sets. \textbf{a,b,c,d,e,f,g,} Observations for the Temperature (\textbf{a}), Star (\textbf{b}), Sunspot (\textbf{c}), Milk (\textbf{d}), Gasoline (\textbf{e}), Car (\textbf{f}), and Lake (\textbf{g}) datasets.}
    \label{fig:probs}
\end{figure}

Each forecasting problem was limited to the first 150 data points when there were more points, and then the observations were subdivided into sets for training, validation, and test in a proportion of 50\%, 25\%, and 25\%, respectively. Fig.~\ref{fig:probs} presents the observations of each problem, with vertical dashed lines separating the training, validation, and test subsets.

As can be seen in terms of the main differences between the problems, Figs.~\ref{fig:temperature-prob}, \ref{fig:star-prob}, and \ref{fig:sunspot-prob} are distinct cases of no trend, while Figs.~\ref{fig:milk-prob}, \ref{fig:gasoline-prob}, and \ref{fig:car-prob} are distinct examples with a constant trend, and finally Fig.~\ref{fig:lake-prob} comprises distinct trends over time. Note that, when converting each subset of observations to a supervised task that arranges consecutive points as inputs to require the next point as output, the last observations of the training set are coupled to the validation set. In the same way, the last observations of the validation set are coupled to the test set. In turn, outputs can be produced only after the first observations of the training set.

\bibliographystyle{ieeetr}
\bibliography{bibliography}

\appendix

\section{Supplementary information}

\subsection{Definition and illustration of the proposed hybridization}

Mathematically defining the sequential combination of quantum and classical models proposed in this work, the first step is to provide an approximate value $\hat{y}_{t+1}^{Q}$ for the next observation using a quantum model Q that takes as input the $n_Q$ preceding observations $y_i$ of the time series. Therefore, the quantum model generates a prediction in the following way:

\begin{equation*}
y_{ \{ t - (n_Q - 1) \} },\ y_{ \{ t - (n_Q - 2) \} },\ \cdots\ ,\ y_{t-1},\ y_t \quad \xmapsto{\qquad \text{Q} \qquad} \quad \hat{y}_{t+1}^{Q}.
\end{equation*}

Previous quantum predictions $\hat{y}_i^Q$ can be subtracted from the corresponding values $y_i$ to generate previous quantum errors $e_i$ as follows, where $i < t+1$:

\begin{equation*}
e_i = y_i - \hat{y}_i^Q.
\end{equation*}

Then, a temporal window composed of the last $n_C$ quantum errors is passed to a classical model C that predicts $\hat{e}_{t+1}^{C}$ as the next error to be made by the quantum model:

\begin{equation*}
e_{ \{ t - (n_C - 1) \} },\ e_{ \{ t - (n_C - 2) \} },\ \cdots\ ,\ e_{t-1},\ e_t \quad \xmapsto{\qquad \text{C} \qquad} \quad \hat{e}_{t+1}^{C}.
\end{equation*}

At this stage, the classical error prediction $\hat{e}_{t+1}^{C}$ can be used to correct the initial approximation $\hat{y}_{t+1}^{Q}$ given by the quantum model. Here, a hybrid prediction $\hat{y}_{t+1}^{Q+C}$ for the observation $y_{t+1}$ is finally obtained by combining quantum and classical predictions in this manner:

\begin{equation*}
\hat{y}_{t+1}^{Q+C} = \hat{y}_{t+1}^{Q} + \hat{e}_{t+1}^{C}.
\end{equation*}

\begin{figure}[htp!]
    \centering
    \includegraphics[width=\textwidth]{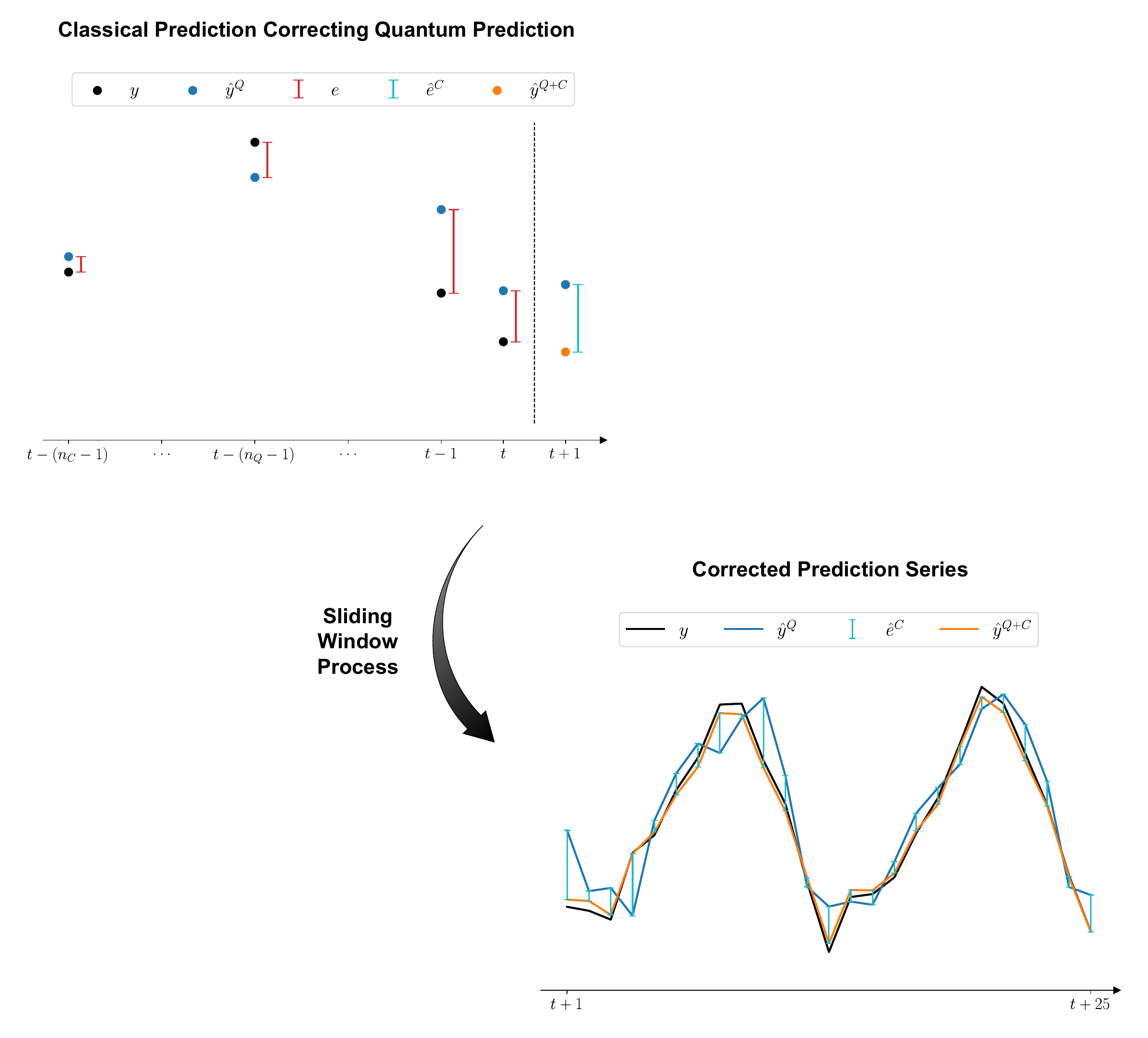}
    \caption{Illustration of the proposed error-correction-based quantum-classical hybridization for time series forecasting. To correct a quantum prediction, first, previous observations $y_i$ and quantum predictions $\hat{y}_i^Q$ are required to determine $n_C$ previous quantum errors $e_i$ that are given as input for a classical model to predict the next error $\hat{e}_{t+1}^{C}$, which are represented by black circles, blue circles, red bars, and a cyan bar, respectively. This classical prediction can then be used to shift the quantum prediction $\hat{y}_{t+1}^{Q}$, also shown as a blue circle, which comes from the last $n_Q$ observations. Finally, the orange circle represents the error-corrected hybrid prediction $\hat{y}_{t+1}^{Q+C}$ for the next observation. Sliding the input windows for the quantum and classical models allows successive corrections over time, that is, the blue line can be point-to-point shifted by the cyan bars. As a result, the hybrid prediction series better approximates the target time series, represented by the orange and black lines, respectively.}
    \label{fig:prop-model-illus}
\end{figure}

Fig.~\ref{fig:prop-model-illus} illustrates all these elements involved in the proposed quantum-classical hybridization based on error correction, in addition to exemplifying an error-corrected quantum prediction series that is obtained through a sliding window process. First, the last observations $y_i$ and quantum predictions $\hat{y}_i^Q$ are required to derive the previous quantum errors $e_i$, which are represented by black circles, blue circles, and red bars, respectively. From the $n_C$ previous quantum errors, a classical model predicts the next error $\hat{e}_{t+1}^{C}$, represented by the cyan bar. The classical prediction can then correct the quantum prediction $\hat{y}_{t+1}^{Q}$, also shown as a blue circle, which comes from the last $n_Q$ observations of the time series. Adding $\hat{e}_{t+1}^{C}$ to $\hat{y}_{t+1}^{Q}$ means shifting the blue circle by the cyan bar, finally giving the orange circle that represents the error-correction-based hybrid prediction $\hat{y}_{t+1}^{Q+C}$ for the next observation in the series.

In addition to being corrected by the classical prediction $\hat{e}_{t+1}^{C}$, the quantum prediction $\hat{y}_{t+1}^{Q}$ is also used to derive the quantum error $e_{t+1}$ as the deviation from the true value $y_{t+1}$. In this way, at the same time that the input window for the quantum model can slide to include $y_{t+1}$, the input window for the classical model can slide to include $e_{t+1}$, and thus $\hat{y}_{t+2}^{Q}$, $\hat{e}_{t+2}^{C}$, and $\hat{y}_{t+2}^{Q+C}$ can be generated. Fig.~\ref{fig:prop-model-illus} also illustrates a hybrid prediction series obtained through this sliding window process for a certain period of time. Such a hybrid prediction series, which is represented by the orange line, results from consecutive shifts of the quantum predictions by the classical predictions, represented by the blue line and the cyan bars, respectively. This correction process is supposed to generate a prediction series that better follows a target time series shown in black. Hence, the proposed hybridization can be seen as a first quantum approximation that is followed by a classical refinement of the predictions.
\subsection{Error metrics and the disagreement in the Sunspot problem}

Three errors metrics were used to assess the quality of model predictions: MSE, MAE, and MAPE. These metrics actually quantify the proximity to the target time series through a central value of prediction errors. Given an $n$-items time series $y = \{ y_0, \cdots, y_{n-1} \}$ and a corresponding prediction series $\hat{y} = \{ \hat{y}_0, \cdots, \hat{y}_{n-1} \}$, the metrics calculate a mean value for the deviations $e = \{ e_0, \cdots, e_{n-1} \}$ in the following ways, where $e_i = y_i - \hat{y}_i$:
\begin{align*}
    \text{MSE} &= \frac{1}{n} \sum_{i=0}^{n-1} e_i^2, \\[10pt]
    \text{MAE} &= \frac{1}{n} \sum_{i=0}^{n-1} |e_i|, \text{ and} \\[10pt]
    \text{MAPE} &= \frac{1}{n} \sum_{i=0}^{n-1} \left| \frac{e_i}{y_i} \right|.
\end{align*}

\begin{figure}[htp!]
    \centering
    \includegraphics[width=0.75\textwidth]{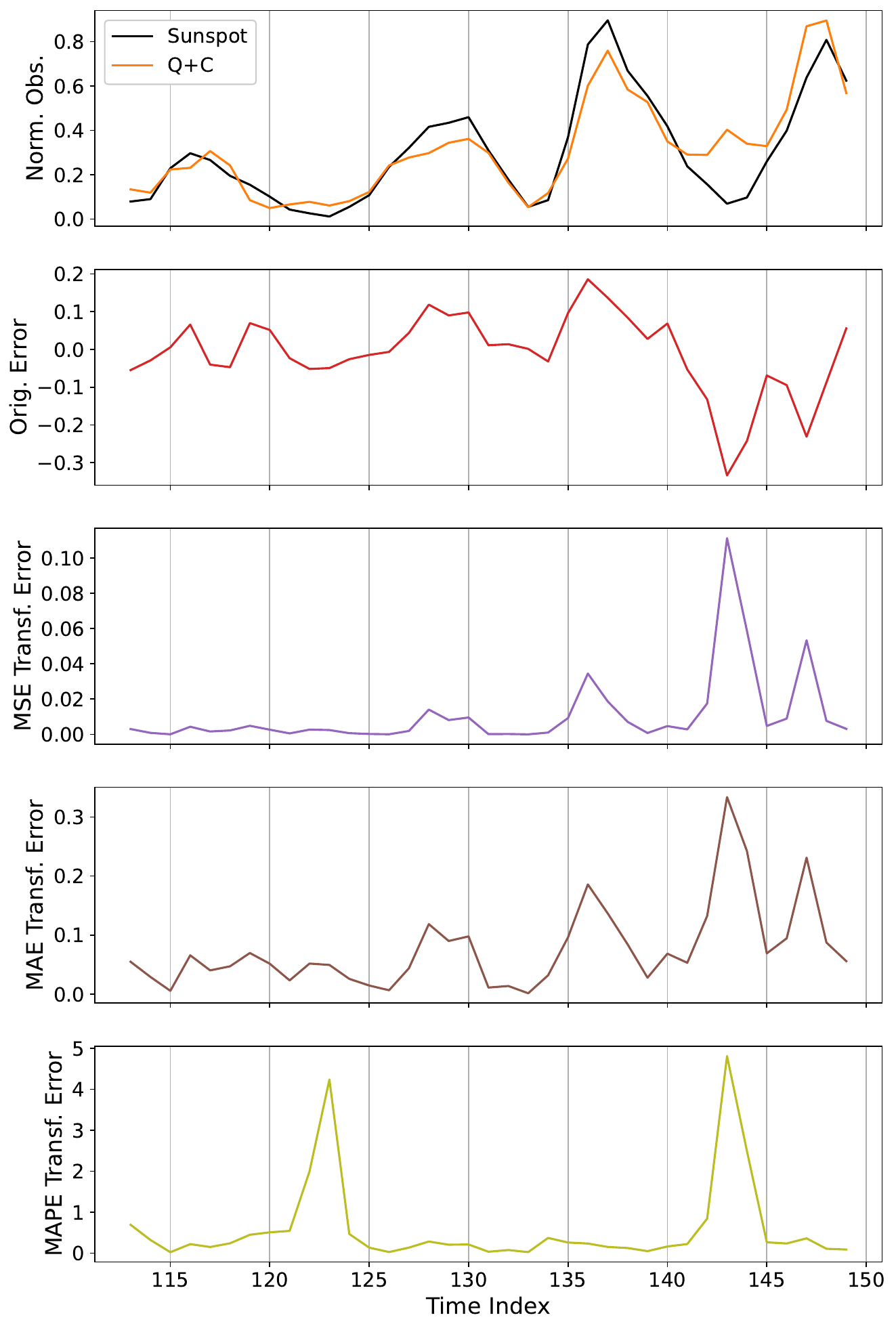}
    \caption{Paired error series that are generated from the original prediction errors of the Q+C model for the Sunspot problem when calculating the MSE, MAE, and MAPE metrics. The first graph contrasts the true observations with the predicted values, represented respectively by the black and orange lines, whose deviations form the original error series of the second graph. Then, the metric calculations transform the original series of errors into the error series of the last three graphs.}
    \label{fig:transf-errs}
\end{figure}

Thus, each metric calculation can be seen as the mean value of a different error series $e^* = \{ e_0^*, \cdots, e_{n-1}^* \}$, where $e_i^*$ is the original $e_i$ transformed by a function $f(\cdot)$. This function is $f(e_i) = e_i^2$ for the MSE metric and $f(e_i) = |e_i|$ for the MAE metric. In turn, the function becomes $f(e_i, y_i) = \left| \frac{e_i}{y_i} \right|$ in the calculation of MAPE. To illustrate such a difference, Fig.~\ref{fig:transf-errs} pairs the three transformed error series that can be generated from the original error series of the Q+C predictions for the Sunspot problem. First, Fig.~\ref{fig:transf-errs} shows the target time series and the model predictions using the black and orange lines, respectively. Then, the original errors are shown, followed by the three transformed error series that arise in the calculation of MSE, MAE, and MAPE, respectively.

\begin{figure}[htp!]
    \centering
    \includegraphics[width=0.75\textwidth]{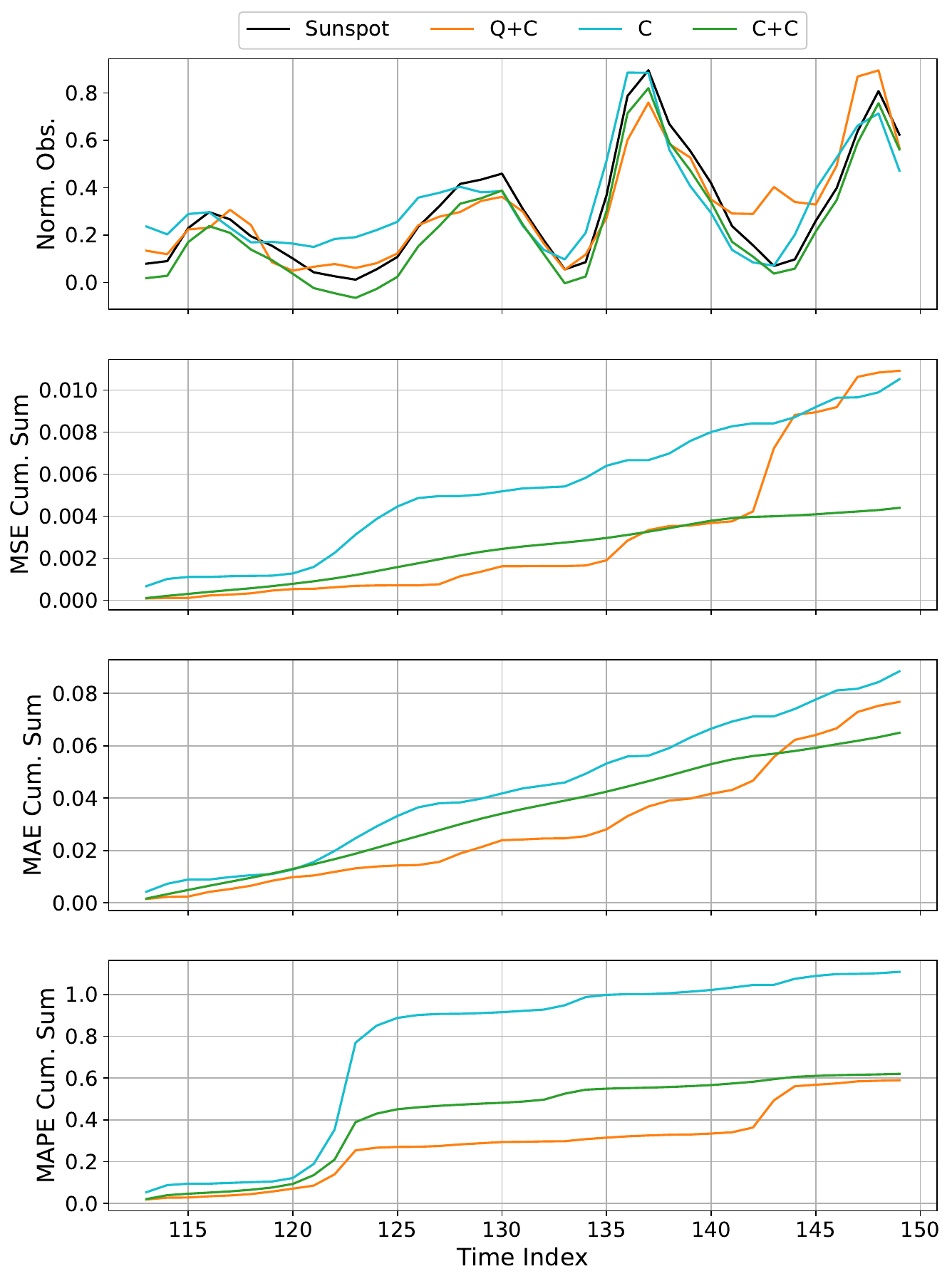}
    \caption{Accumulation of errors over time in the perspective of the MSE, MAE, and MAPE metrics for the Q+C, C, and C+C models in the Sunspot problem. The first graph contrasts the target time series with the three model predictions, represented by the black, orange, cyan, and green lines, respectively. Then, three graphs pair the impact of each original deviation into the error accumulation that produces the final mean values of each metric.}
    \label{fig:sum-errs}
\end{figure}

Due to the squaring operation, the MSE transformation produced a new series where errors near zero were attenuated and large errors were amplified. On the other hand, the MAE transformation preserves the error magnitudes, only reflecting negative errors across zero. Finally, in the MAPE metric, the errors made in the peaks were attenuated, while the errors made in the valleys were amplified, since the transformation divides each error by the corresponding target value. Therefore, the consequence of each error depends on the metric, which can even cause divergence when concluding about the quality of multiple models using different metrics. Fig.~\ref{fig:sum-errs} shows each metric estimation as a cumulative sum of errors $\frac{f(e_i)}{n}$ over time for the predictions of the Q+C, C, and C+C models in Sunspot through the orange, cyan, and green lines.

Basically, the C and C+C models accumulated errors at a constant growth rate until forming the final mean value of the MSE metric, although the rate for the C model is higher. In this case, the Q+C model accumulated the lowest errors until the last valley of the target series, where the errors begin to impose a growth rate that even made Q+C the worst model in the end. A similar behavior occurred for all models when the errors were accumulated in view of MAE. However, the growth rate that emerged around the last valley for the Q+C model was lower this time, in a way that Q+C surpassed the error accumulation of the C+C model only. The MAPE metric, in turn, penalized with sharp growths only the errors in valleys, which allowed the Q+C model to preserve the overall lowest accumulation of errors, followed by the C+C and C models. Thus, the different penalty mechanisms can make the error metrics really disagree in the end about the quality of the models, as clarified here.
\subsection{Results for all single and hybrid models applied in the problems}

\begin{figure}[htp!]
    \centering
    
    \begin{subfigure}{0.48\textwidth}
        \includegraphics[width=\textwidth]{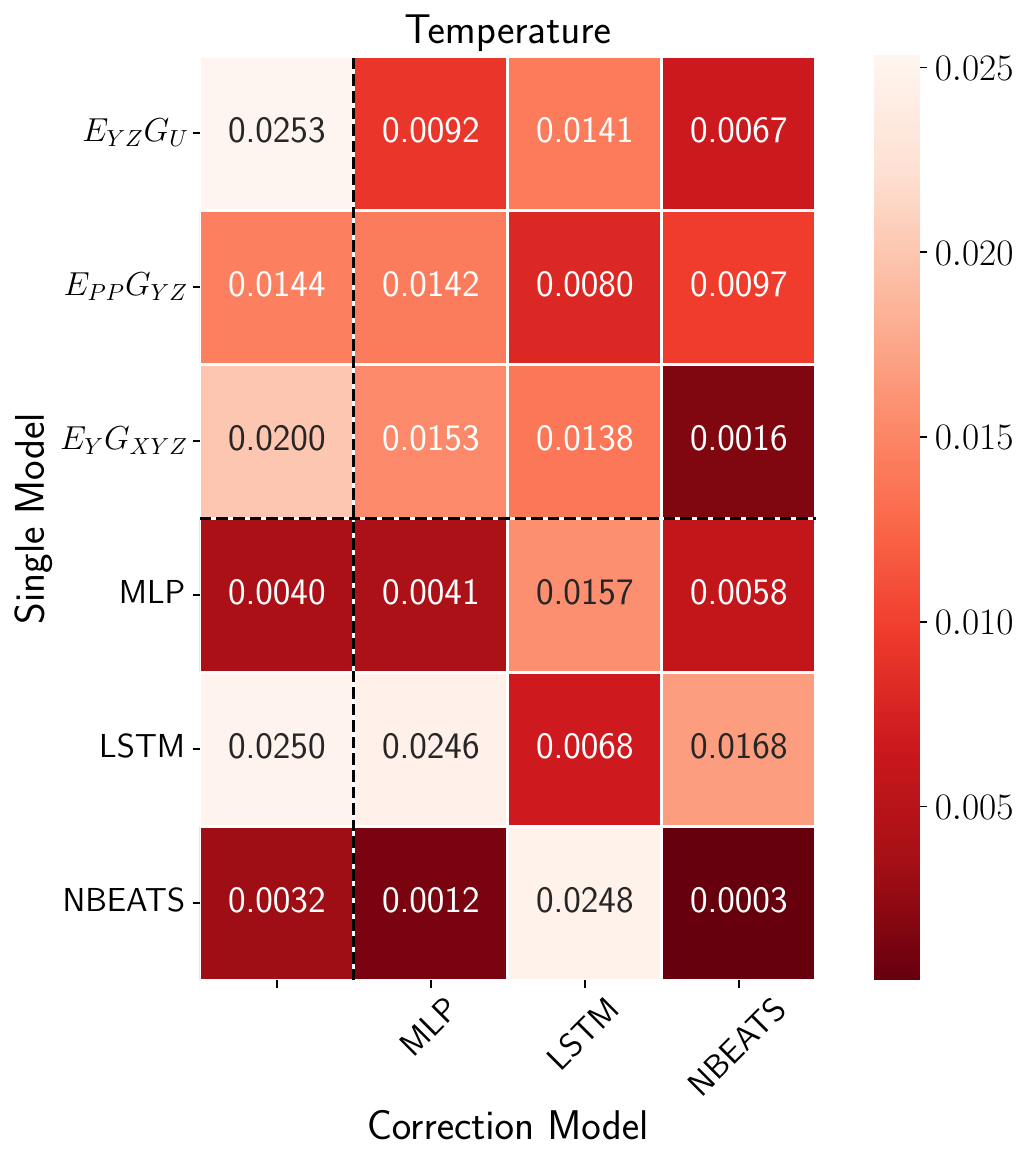}
        \caption{}
        \label{fig:temperature-hm}
    \end{subfigure}
    \hfill
    \begin{subfigure}{0.48\textwidth}
        \includegraphics[width=\textwidth]{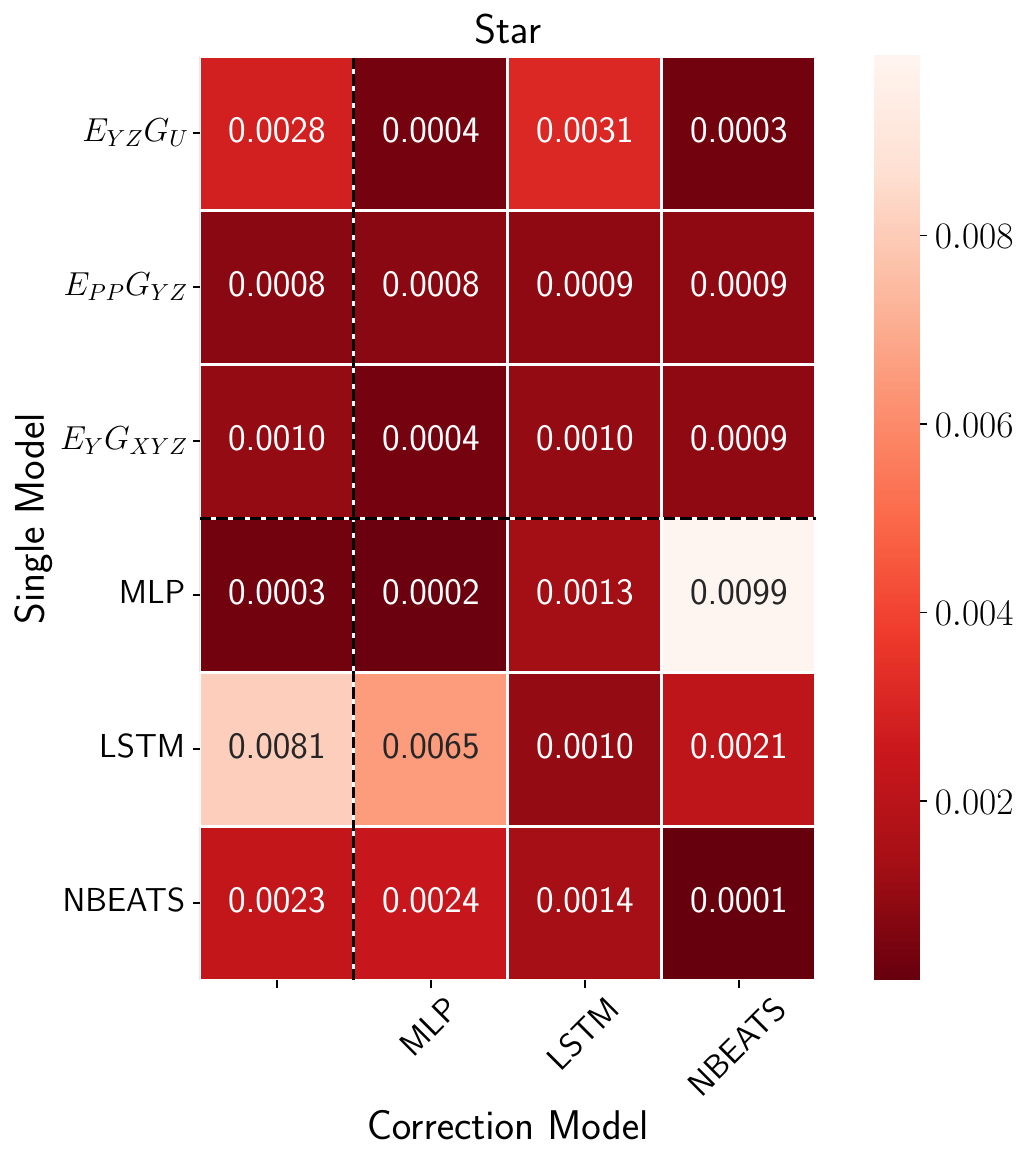}
        \caption{}
        \label{fig:star-hm}
    \end{subfigure}

    \begin{subfigure}{0.48\textwidth}
        \includegraphics[width=\textwidth]{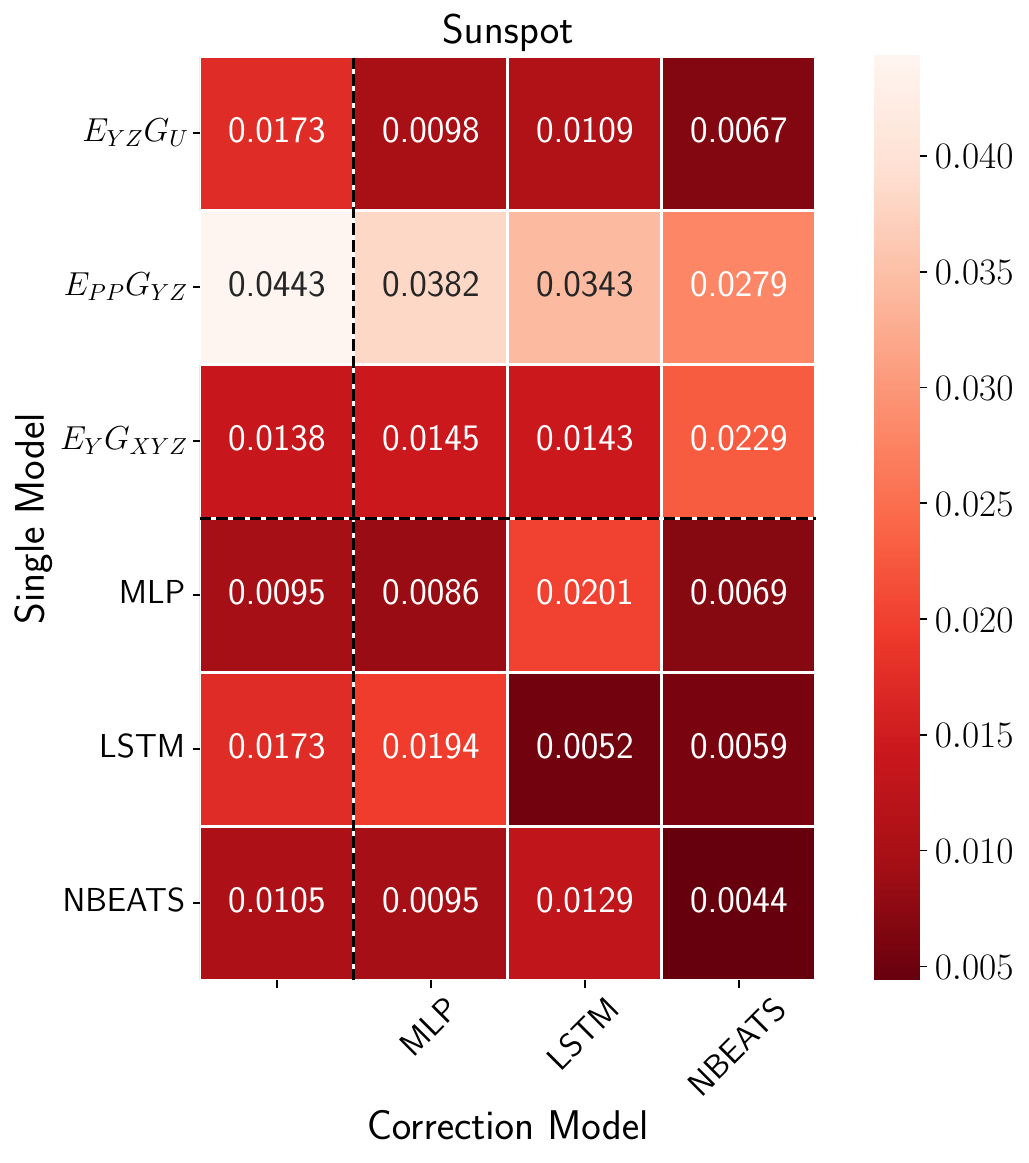}
        \caption{}
        \label{fig:sunspot-hm}
    \end{subfigure}
    \hfill
    \begin{subfigure}{0.48\textwidth}
        \includegraphics[width=\textwidth]{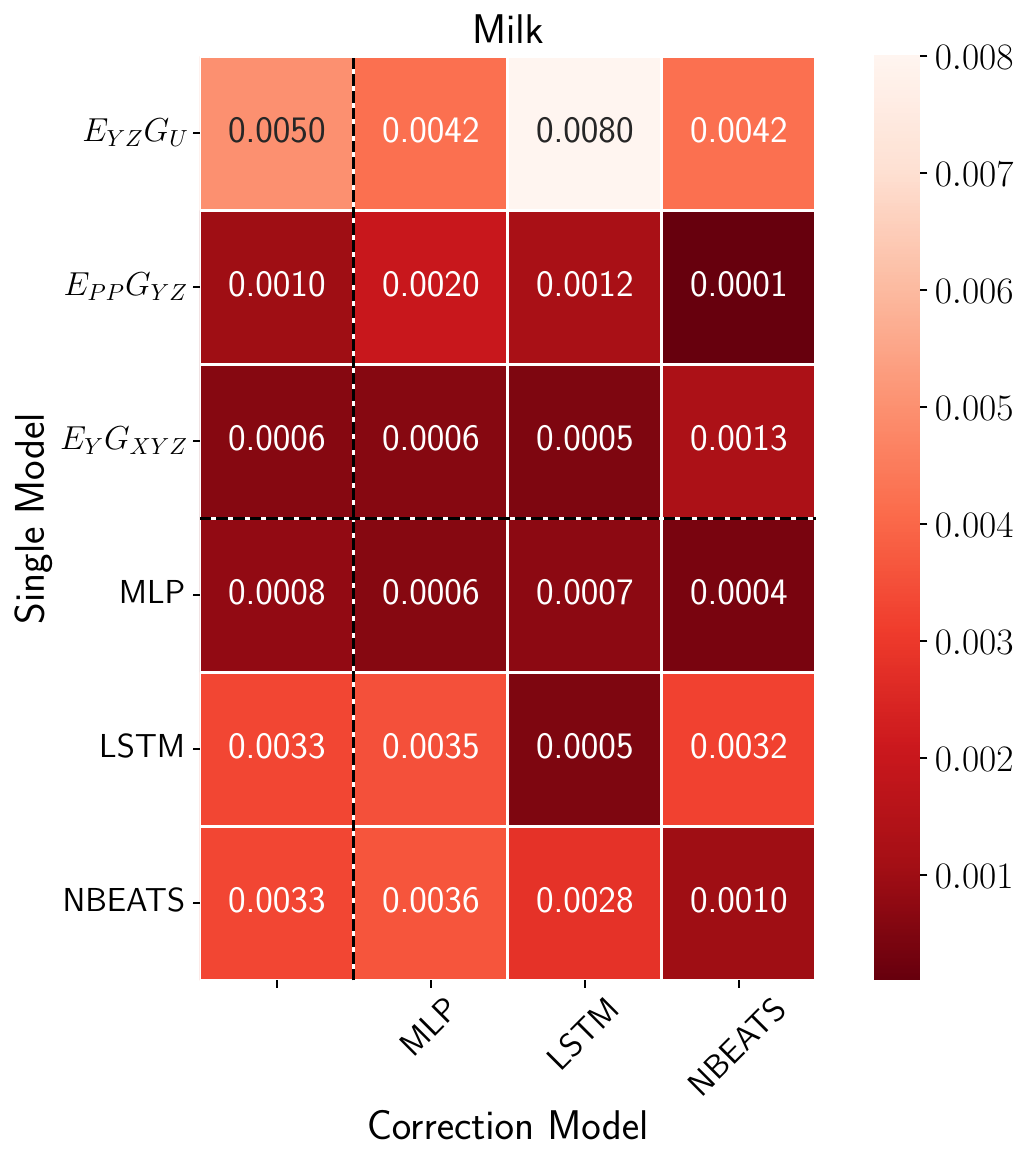}
        \caption{}
        \label{fig:milk-hm}
    \end{subfigure}

    \caption{Heat maps with the MSE values of all single and hybrid models in the first four of the seven problems. The color bars depend on the minimum and maximum MSE values obtained in each problem. The models are arranged in a way that the applied single models are in an axis and the applied correction models are in the other axis. Black dashed lines still separate each class of single and hybrid models. \textbf{a,b,c,d,} Heat maps for the Temperature (\textbf{a}), Star (\textbf{b}), Sunspot (\textbf{c}), and Milk (\textbf{d}) datasets.}
    \label{fig:part1-hms}
\end{figure}

\begin{figure}[htp!]
    \centering
    
    \begin{subfigure}{0.48\textwidth}
        \includegraphics[width=\textwidth]{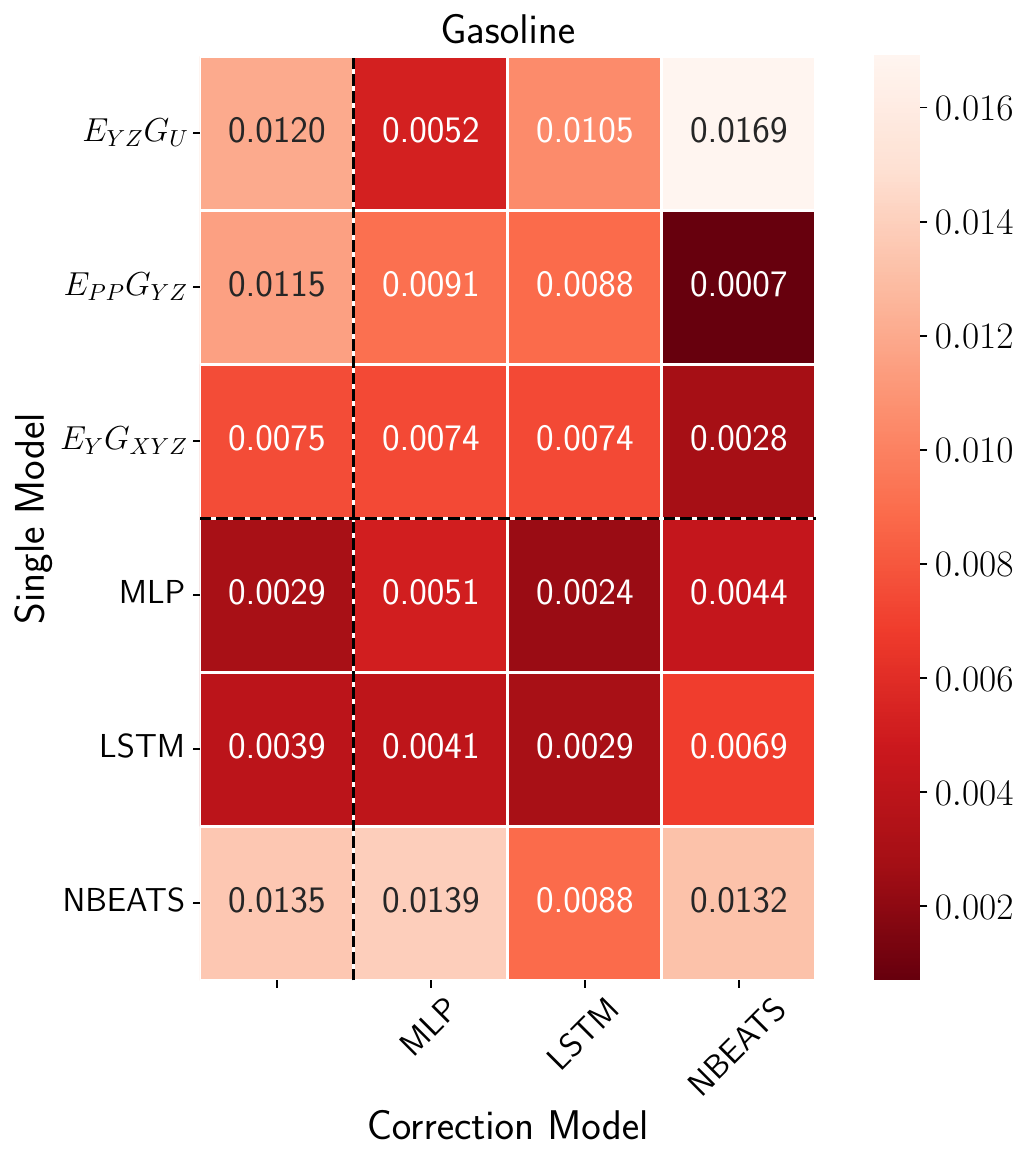}
        \caption{}
        \label{fig:gasoline-hm}
    \end{subfigure}
    \hfill
    \begin{subfigure}{0.48\textwidth}
        \includegraphics[width=\textwidth]{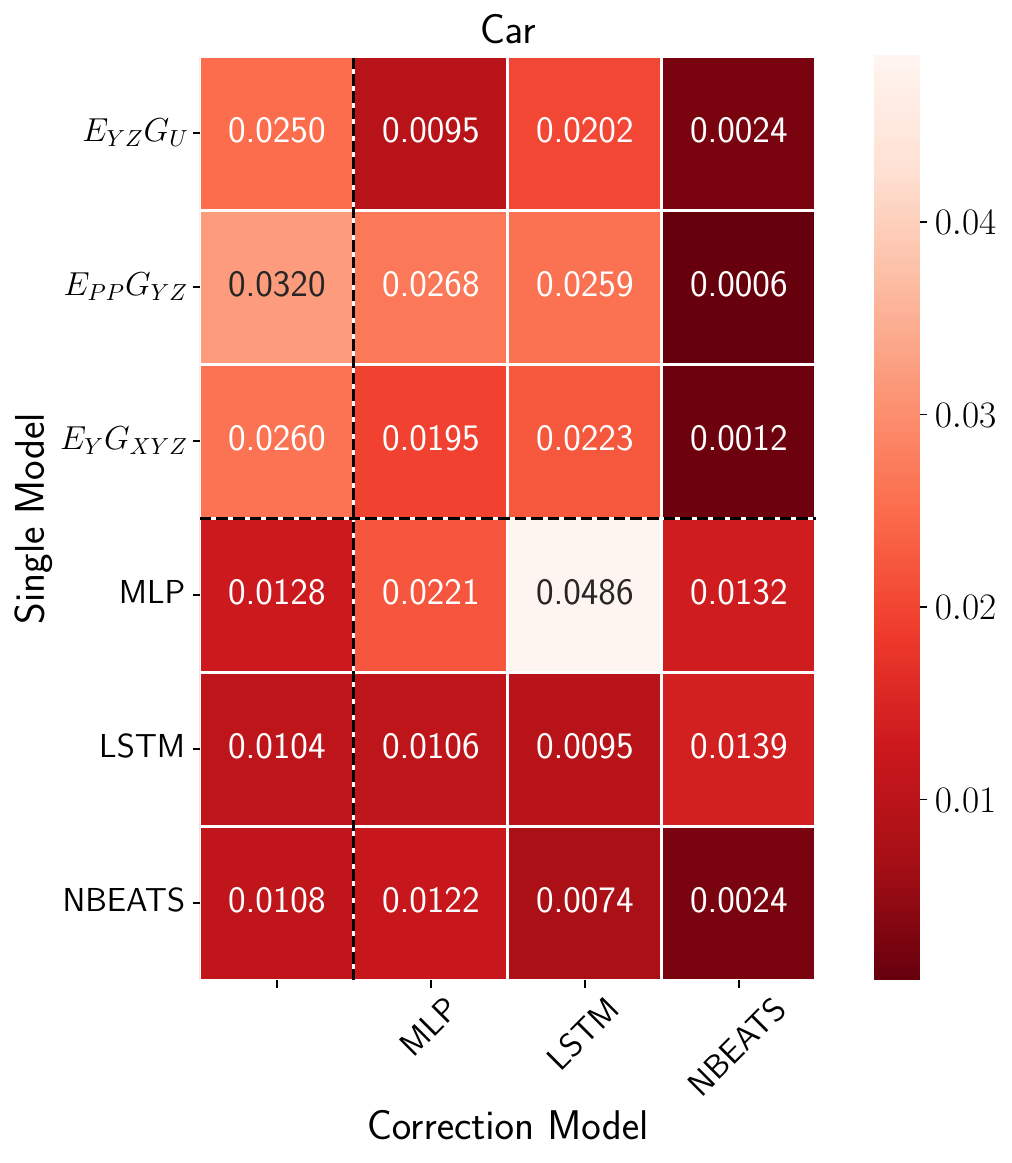}
        \caption{}
        \label{fig:car-hm}
    \end{subfigure}

    \begin{subfigure}{0.48\textwidth}
        \includegraphics[width=\textwidth]{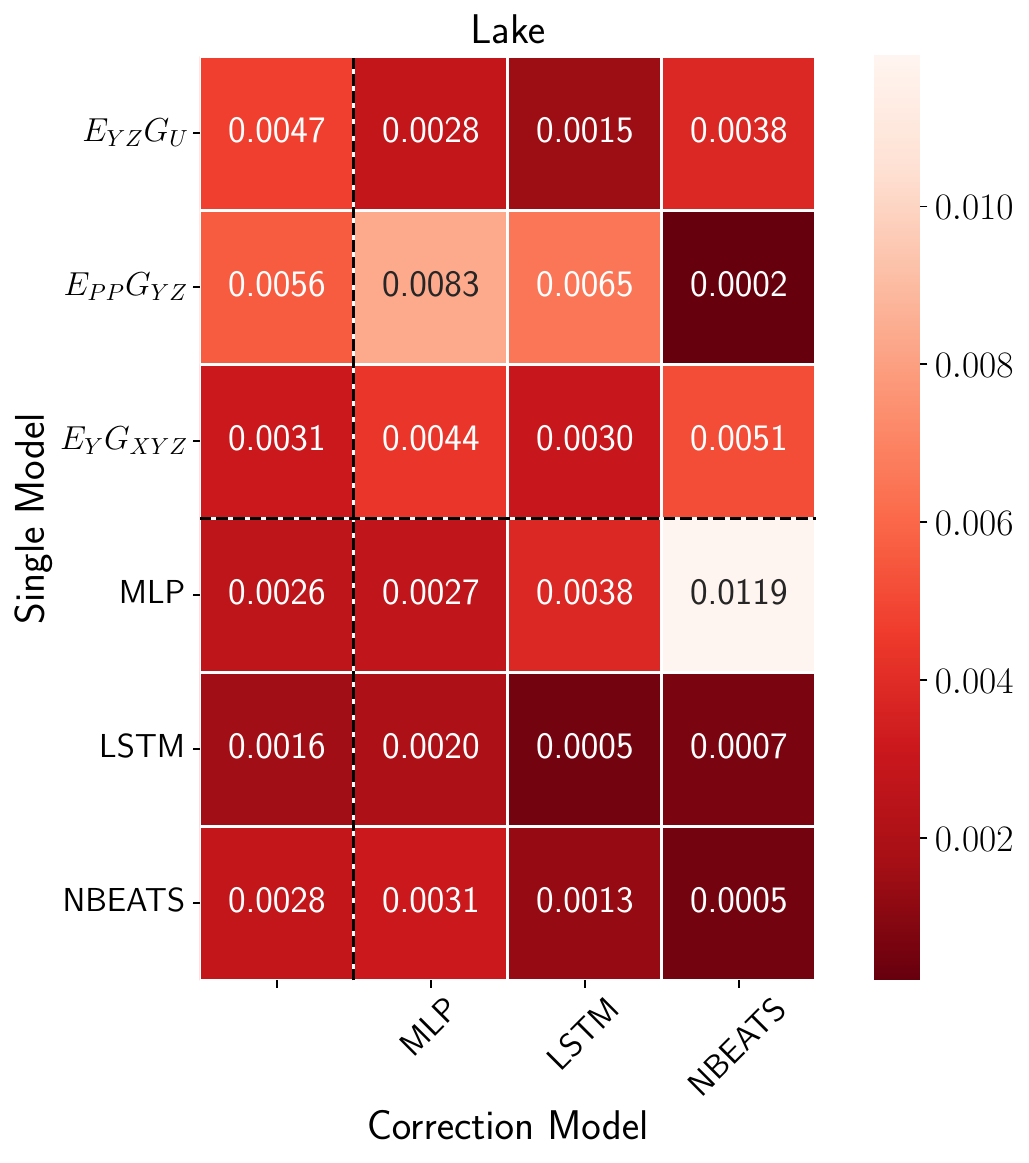}
        \caption{}
        \label{fig:lake-hm}
    \end{subfigure}
    
    \caption{Heat maps with the MSE values of all single and hybrid models in the last three of the seven problems. The color bars depend on the minimum and maximum MSE values obtained in each problem. The models are arranged in a way that the applied single models are in an axis and the applied correction models are in the other axis. Black dashed lines still separate each class of single and hybrid models. \textbf{a,b,c,} Heat maps for the Gasoline (\textbf{a}), Car (\textbf{b}), and Lake (\textbf{c}) datasets.}
    \label{fig:part2-hms}
\end{figure}

The main results already shown in this work considered the best model within each class of models: classical single models, quantum-classical hybrid models, and classical-classical hybrid models. Here, a lower level of abstraction is achieved by presenting the results for all models applied in the experiments. Figs.~\ref{fig:part1-hms} and~\ref{fig:part2-hms} show the MSE values obtained by the models in the unseen data of each problem. By using heat maps that vary with the minimum and maximum MSE values obtained in each problem, the performance of the models can be visually contrasted, especially identifying the regions of intense heat per problem. The lower the MSE value, the better, and therefore the darker the red in the heat map. Each heat map compresses the applied models by arranging the single models in an axis and the correction models in the other axis, where the first column represents no models correcting the single models. The black dashed lines still separate the quantum single models, the classical single models, the quantum-classical hybrid models, and the classical-classical hybrid models.

As shown in Fig.~\ref{fig:temperature-hm}, the $E_{PP}G_{YZ}$ model was the best among the quantum single models in Temperature, although the MLP and NBEATS models performed considerably better. In this problem, all quantum models were able to produce a moderate to high performance when corrected by a classical model, especially the combination $E_YG_{XYZ}$+NBEATS that outperformed the classical single models and competed strongly with the best classical-classical hybrid models. The best classical-classical models in Temperature were NBEATS+MLP and NBEATS+NBEATS. Fig.~\ref{fig:star-hm} shows the $E_{PP}G_{YZ}$ and $E_YG_{XYZ}$ models retaining the best performance as quantum single models, underperforming only the MLP model between the classical single models in Star. Generally, the quantum-classical hybrid models produced moderately the same results as the best quantum single models in Star. However, the best quantum-classical models, which were $E_{YZ}G_U$+MLP, $E_{YZ}G_U$+NBEATS, and $E_YG_{XYZ}$+MLP, competed strongly with both the MLP model and the best classical-classical models, which were MLP+MLP and NBEATS+NBEATS.

In the Sunspot problem, the best quantum single model was the $E_YG_{XYZ}$ model, although classical single models such as MLP and NBEATS performed better, as shown in Fig.~\ref{fig:sunspot-hm}. When combining quantum and classical models in this problem, only the $E_{YZ}G_U$+NBEATS model really produced a better result, even outperforming the classical single models and competing with the best classical-classical models. The LSTM+LSTM and all classical-classical combinations with NBEATS as the correction model were the best classical-classical models in Sunspot. On the other hand, the $E_YG_{XYZ}$ model excelled in predicting the Milk problem, as shown in Fig.~\ref{fig:milk-hm}, since this quantum single model competed with the MLP model as the best classical single model and with the MLP+NBEATS and LSTM+LSTM models as the best classical-classical models. Furthermore, when considering the quantum-classical models in Milk, the $E_{PP}G_{YZ}$+NBEATS model excelled even more, producing the best overall result between all models in this problem.

When the quantum single models were applied in the Gasoline, Car, and Lake problems, $E_YG_{XYZ}$ was an outstanding model again, as shown in Figs.~\ref{fig:gasoline-hm}, \ref{fig:car-hm}, and \ref{fig:lake-hm}. Specifically in Car, $E_YG_{XYZ}$ shared the position of the best quantum model with $E_{YZ}G_U$. In these three problems, except for NBEATS in Gasoline, all classical single models performed better than the best quantum single model. However, the $E_{PP}G_{YZ}$+NBEATS model, which was the best quantum-classical hybrid model in the three problems, outperformed the best classical single models and even the best classical-classical hybrid models. The best classical single models in Gasoline and Lake were, respectively, the MLP and LSTM models. In the Car problem, the three classical single models presented a comparable performance. Regarding the classical-classical models, the best models were MLP+LSTM in Gasoline, NBEATS+NBEATS in Car, and both LSTM+LSTM and NBEATS+NBEATS in Lake.

In summary, the best quantum single models were unable to capture time patterns as the best classical single models, and then performed worse. Among the single models, the simpler architectures prevailed, since $E_{Y}G_{XYZ}$ was the best quantum model and MLP was the best classical model in most cases. The $E_{Y}G_{XYZ}$ model explores the simplest data encoding among the applied quantum models, at the same time that $E_{Y}G_{XYZ}$ adjusts fewer parameters per learnable layer than $E_{PP}G_{YZ}$ and explores a simpler structure of $CX$ gates than $E_{YZ}G_U$. On the other hand, the $E_{Y}G_{XYZ}$ and MLP models rarely participated in the best quantum-classical hybrid models.

Instead, the other quantum models were mostly required, especially $E_{PP}G_{YZ}$, which explores a complex quantum encoding to then capture time patterns that ultimately complement the classical models applied in the sequence. To effectively capture complementary patterns after the application of a quantum model, the strength of a deep neural network such as NBEATS was regularly required in the best quantum-classical models. The combination of more complex architectures in the quantum and classical models finally developed a novel forecasting capacity. Such an augmented quantum-classical capacity was able to overcome even the capacity of classical-classical hybrid models, even though the best classical-classical models repeatedly leveraged the NBEATS model as the first applied model, the correction model, and, mainly, both.

\end{document}